\documentclass[12pt,onecolumn,draftclsnofoot,peerreview]{IEEEtran}
\usepackage[T1]{fontenc}
\usepackage[latin9]{inputenc}
\usepackage{units}
\usepackage{amsmath}
\usepackage{amsthm}
\usepackage{amssymb}
\usepackage{graphicx}
\usepackage{setspace}
\usepackage[unicode=true,
 bookmarks=true,bookmarksnumbered=true,bookmarksopen=true,bookmarksopenlevel=1,
 breaklinks=false,pdfborder={0 0 0},pdfborderstyle={},backref=false,colorlinks=false]
 {hyperref}
\hypersetup{pdftitle={Your Title},
 pdfauthor={Your Name},
 pdfpagelayout=OneColumn, pdfnewwindow=true, pdfstartview=XYZ, plainpages=false}

\makeatletter

\providecommand{\tabularnewline}{\\}

 \let\oldforeign@language\foreign@language
 \DeclareRobustCommand{\foreign@language}[1]{%
   \lowercase{\oldforeign@language{#1}}}
\theoremstyle{plain}
\newtheorem{thm}{\protect\theoremname}
\theoremstyle{plain}
\newtheorem{lem}[thm]{\protect\lemmaname}

\usepackage[caption=false,font=footnotesize]{subfig}

\@ifundefined{showcaptionsetup}{}{%
 \PassOptionsToPackage{caption=false}{subfig}}
\usepackage{subfig}
\makeatother

\providecommand{\lemmaname}{Lemma}
\providecommand{\theoremname}{Theorem}

\begin{document}

\title{Coverage and Handoff Analysis of 5G Fractal Small Cell Networks}

\author{Jiaqi~Chen,~\IEEEmembership{Student~Member,~IEEE,} Xiaohu~Ge,~\IEEEmembership{Senior~Member,~IEEE,}
and Qiang Ni,~\IEEEmembership{Senior~Member,~IEEE}\thanks{Jiaqi~Chen, Xiaohu~Ge (Corresponding author) are with School of
Electronic Information and Communications, Huazhong University of
Science and Technology, Wuhan 430074, Hubei, P. R. China, e-mail:
\protect\href{mailto:chenjq_hust@mail.hust.edu.cn}{chenjq\_{}hust@mail.hust.edu.cn},
\protect\href{mailto:xhge@mail.hust.edu.cn}{xhge@mail.hust.edu.cn}.}\thanks{Qiang~Ni is with InfoLab21, School of Computing and Communications,
Lancaster University, Lancaster, LA1 4WA, UK, e-mail: \protect\href{mailto:q.ni@lancaster.ac.uk}{q.ni@lancaster.ac.uk}.}}

\markboth{IEEE transactions on Wireless Communications}{Jiaqi Chen \MakeLowercase{\emph{et al.}}: Coverage and Handoff
Analysis of 5G Fractal Small Cell Networks}
\maketitle
\begin{abstract}
It is anticipated that much higher network capacity will be achieved
by the fifth generation (5G) small cell networks incorporated with
the millimeter wave (mmWave) technology. However, mmWave signals are
more sensitive to blockages than signals in lower frequency bands,
which highlights the effect of anisotropic path loss in network coverage.
According to the fractal characteristics of cellular coverage, a multi-directional
path loss model is proposed for 5G small cell networks, where different
directions are subject to different path loss exponents. Furthermore,
the coverage probability, association probability, and the handoff
probability are derived for 5G fractal small cell networks based on
the proposed multi-directional path loss model. Numerical results
indicate that the coverage probability with the multi-directional
path loss model is less than that with the isotropic path loss model,
and the association probability with long link distance, \textit{e.g.},
150m, increases obviously with the increase of the effect of anisotropic
path loss in 5G fractal small cell networks. Moreover, it is observed
that the anisotropic propagation environment is having a profound
impact on the handoff performance. Meanwhile, we could conclude that
the resulting heavy handoff overhead is emerging as a new challenge
for 5G fractal small cell networks.
\end{abstract}


\IEEEpeerreviewmaketitle{}

\section{Introduction}

\IEEEPARstart{O}{ne} of the primary goals of the fifth generation
(5G) mobile communication is to provide 100Mbps to 1Gbps data rates
anytime and anywhere. This goal can be achieved by combining a variety
of 5G new technologies, such as massive multi-input multi-output (MIMO),
millimeter wave (mmWave), and small cell networks \cite{key-1,key-71}. Base
stations (BSs) equipped with large antenna arrays could achieve directivity
gains through precoding\slash{}beamforming and could transmit mmWave
signals, which results in high data rates. For example, the Samsung\textquoteright s
prototype \cite{key-3}, which is a 4$\times$8 uniform planar array
working at 27.925GHz, is able to offer excellent performance, where
the reported peak data rate with no mobility is about 1Gbps over a
range up to 1.7 kilometers in Line-of-Sight (LoS) transmissions or
200 meters in Non-Line-of-Sight (NLoS) transmissions. The mmWave signals
are more sensitive to blockages than signals in lower frequency bands
due to the shorter wavelength \cite{key-4,key-5}. Thus, the link
distance is reduced when the mmWave signals are transmitted, which
results in the trend of small cells \cite{key-55,key-56}. In small
cell networks, users frequently switch from a small cell BS (SBS)
to another SBS since the coverage area of a small cell is significantly
smaller than that of a macro cell. The frequent handoff issue in a
small cell network with mmWave system becomes an essential factor
that affects the network performance. On the other hand, the fractal
coverage of small cells and the propagation environment have the critical
influence on the handoff decision. Thus, it is vital to analyze the
coverage and handoff performance of 5G small cell networks, considering
the anisotropic mmWave signal attenuation in complex urban scenarios.

The coverage performance of small cell networks has been well studied.
In small cell networks, multiple classes of BSs such as macro BSs,
hotspot BSs, and femtocell BSs are incorporated to provide better
coverage for a large number of users. Based on the stochastic geometry
theory, the downlink and uplink coverage probabilities were derived
for multi-tier heterogeneous cellular networks (HCNs) \cite{key-6}.
In the HCN with a dense small cell topology, the coverage probability
for a typical user connecting to the strongest BS signal was analyzed
in \cite{key-14} by incorporating a flexible notion of BS load with
a new idea of conditionally thinning the interference field, where
penetrating analyses were extended to investigate the impact of network
models on the HCN coverage performance. From the point of view of
green communication, the coverage probability for a particular cell
association scheme, such as the maximum received power association
(MRPA) and the nearest base station association (NBA) schemes in the
downlink of an HCN, was derived aimed at establishing the fundamental
limits on the achievable link and network energy efficiencies \cite{key-7-1}.

Moreover, transmission links could be distinguished as LoS vs. NLoS
ones based on the differences in the path loss characteristic caused
by blockages. Such differences become more influential in 5G mmWave
systems than in the conventional network, and are generally taken
into consideration when the performance of the mmWave system is analyzed.
A comprehensive overview of mathematical models and analytical techniques
for mmWave cellular systems were summarized in \cite{key-7}, where
a baseline analytical approach based on stochastic geometry was presented
which allows the computation of statistical distributions of the downlink
signal-to-interference-plus-noise-ratio (SINR). A better representation of the outage possibilities of mmWave communications
was provided in \cite{key-9}, considering realistic path-loss and
blockage models, which were derived from recently reported experimental
data. In \cite{key-9}, simple and exact integrals as well as approximated
and closed-form formulas for computing the coverage probability and
the average rate were obtained. The coverage and rate performance
of mmWave cellular networks in major metropolitan cities like Manhattan
and Chicago were studied in \cite{key-15} to confirm that dense BS
deployment is the key to achieve both better coverage and higher rate
in mmWave cellular networks. Using a distance-dependent LoS probability
function calculated by modeling the blockage distribution as a Poisson
point process (PPP), a general framework was proposed in \cite{key-25}
to evaluate the coverage and rate performance in mmWave cellular networks.
Only two types of transmission links, LoS and NLoS links are considered
in the above analysis. In fact, the path loss exponents of different
transmission links are entirely different due to the anisotropic propagation
environment. In this case, an anisotropic path loss model should be
provided to analyze the coverage performance of small cell networks
with the mmWave system, compared with the distinction between LoS
and NLoS transmission links.

Considering the more irregular coverage of mmWave cellular networks
and smaller coverage area of a small cell, users have to switch from
an SBS to another SBS frequently. An approach for calculating the
handoff rate was proposed in \cite{key-16} based on the angle mobility
model. The handoff rate in mmWave 5G systems was investigated in \cite{key-17},
where the typical average handoff interval was shown to be in the
range of several seconds. The handoff rate for a user was obtained
in \cite{key-18} for an irregular cellular network with the access
point locations modeled as a homogeneous PPP. The vehicular handoff
rate and the vehicular overhead ratio were utilized in \cite{key-19}
to evaluate the distance based vehicular mobility performance in 5G
cooperative small cell networks. Efficient handoff algorithms were
proposed in \cite{key-20,key-21,key-22} to minimize the handoff and
connection failure rates. In the most handoff analysis literature,
the variety of the propagation environment during the user's moving
is ignored in analyzing the handoff performance. However, due to the
anisotropic propagation environment, changes in the path loss exponent
of the transmission links during the user's moving have a significant
influence on the handoff performance, especially in small cell networks
with the mmWave system.

The research efforts in {[}6\textendash 18{]} were conducted with
an assumption that the path loss is isotropic in a cellular scenario,
where the most common path loss model is the uniform path loss model,
\textit{i.e.}, the path loss exponent is a constant over the entire
plane. A piecewise linear path loss model was proposed in \cite{key-27}
where different distance ranges are subject to different path loss
exponents. Furthermore, researchers at New York University built a
novel channel simulation software named NYUSIM based on the statistical
spatial channel model for broadband mmWave wireless communication
systems \cite{key-51,key-52}. In \cite{key-51}, the path loss exponents
of LoS and NLoS transmissions were configured as different constants.
However, our work in \cite{key-23} indicated that the measured wireless
cellular coverage boundary is irregular with respect to directions,
and the statistical fractal characteristic of coverage
boundary was proved to exist widely. Such a fractal characteristic
is illustrated by the spectral density power-law behavior and the
slowly decaying variances in the angle domain \cite{key-23}. Moreover,
in \cite{key-28} the single cell coverage boundary was proved to
bear a visible burst pattern, and the probability density function
(PDF) of the single cell boundary was shown to have the heavy tail
characteristic. In \cite{key-61}, the fractal coverage characteristic
was first used to evaluate the performance of small cell networks.
From the above result, it is evident that the propagation environment
is anisotropic in a mmWave scenario. Although there exist a large
amount of meaningful and essential studies for performance analysis
of the small cell networks with mmWave, the impact of such anisotropic
propagation environments is still not well investigated. Along with the prosperity of machine learning, rich learning schemes \cite{key-72} obviously facilitate the research of the complex and accurate anisotropic path loss models.

In this paper, based on the fractal characteristics of cellular coverage
\cite{key-23}, a multi-directional path loss model is proposed for
5G fractal small cell networks, where different directions are subject
to different path loss exponents. Based on the proposed multi-directional
path loss model, the coverage performance in SINR and rate terms are
analyzed to estimate the impact of the anisotropic propagation environment.
The handoff decision is made based on the association scheme and the
channel state. Some existing analysis for the handoff performance
works based on the assumption that the nearest neighbor SBS is the
one with the highest SINR. However, the association state with the
maximum SINR scheme is more complex in the anisotropic scenario than
in the isotropic scenario. The association probability with the multi-directional
path loss model is investigated to imply the influence of the anisotropic
propagation environment and to lay the foundation for handoff analysis.
Furthermore, the handoff performance of 5G fractal small cell networks
is evaluated by the handoff probability and handoff rate. The handoff
probability implies the performance of an individual handoff operation,
and the handoff rate illustrates the overall network performance.
The main contributions of this paper are as follows:
\begin{enumerate}
\item Based on the fractal characteristics of cellular coverage, a multi-directional
path loss model is proposed to analyze the impact of an anisotropic
propagation environment on the performance of 5G fractal small cell
networks, where different directions are subject to different path
loss exponents.
\item Based on the multi-directional path loss model, the coverage probability
and the association probability concerning the desired link distance
are derived to show that more complex propagation environments have
negative impacts on the coverage performance of the 5G fractal small
cell network. The assumption that the nearest base station for a user
is the one with the highest SINR is typically not supported in 5G
fractal small cell networks.
\item Based on the theoretical analysis on coverage probability and association
probability of 5G fractal small cell networks, the impact of the anisotropic
propagation environment on the handoff performance is first investigated
by analyzing the handoff probability and handoff rate. Simulation
results show that the anisotropic propagation environment has a profound
effect on the handoff performance.
\end{enumerate}
~~The rest of the paper is organized as follows. Section II describes
the system model. The coverage and association probability analyses
are presented in Section III. The handoff probability and handoff
rate are derived in Section IV. The simulation results are presented
in Section V. Finally, conclusions are drawn in Section VI.

\section{System Model}

We consider a downlink small cell network in this paper. Without loss
of generality, SBSs are assumed to be deployed randomly over an infinite
plane $\mathbb{R}^{2}$ and modeled as a homogeneous PPP of intensity
$\lambda$ \cite{key-53}:
\begin{equation}
\Phi=\left\{ x_{i},i=1,2,3,\ldots\right\} ,
\end{equation}
where $x_{i}$ is a Polar coordinate denoting the location of $\textrm{SBS}_{i}$,
\textit{i.e.}, $x_{i}=\left(r_{i},\theta_{i}\right)$, with radial
coordinate $r_{i}$ denoting the distance between $\textrm{SBS}_{i}$
and the coordinate origin, and angular coordinate $\theta_{i}$. A
typical user is assumed to be located at the coordinate origin, denoted
as $u_{0}$.

Considering that the mmWave technology is excepted to be adopted in
5G small cell networks, a multi-directional path loss model is proposed
in this section and aims at analyzing the impact of anisotropic propagation
environment on the performance of mmWave small cell networks. What\textquoteright s
more, the channel model is introduced.

\subsection{Multi-directional Path Loss Model}

\begin{figure}[h]
\subfloat[\label{fig:scenario}A typical anisotropic propagation scenario.]{\includegraphics[width=7cm]{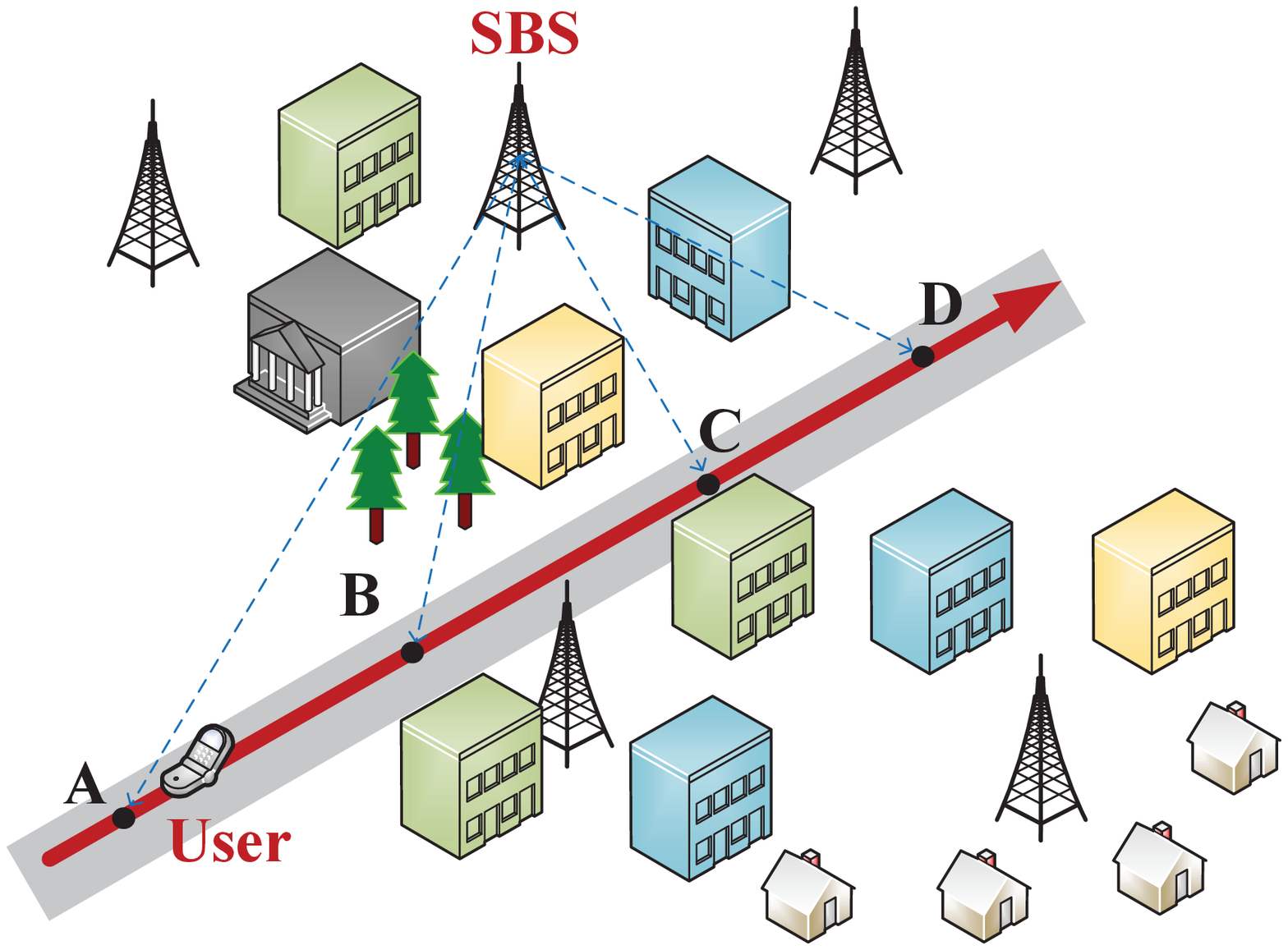}

}\subfloat[\label{fig:path loss model}The multi-directional path loss model.]{\includegraphics[width=7cm]{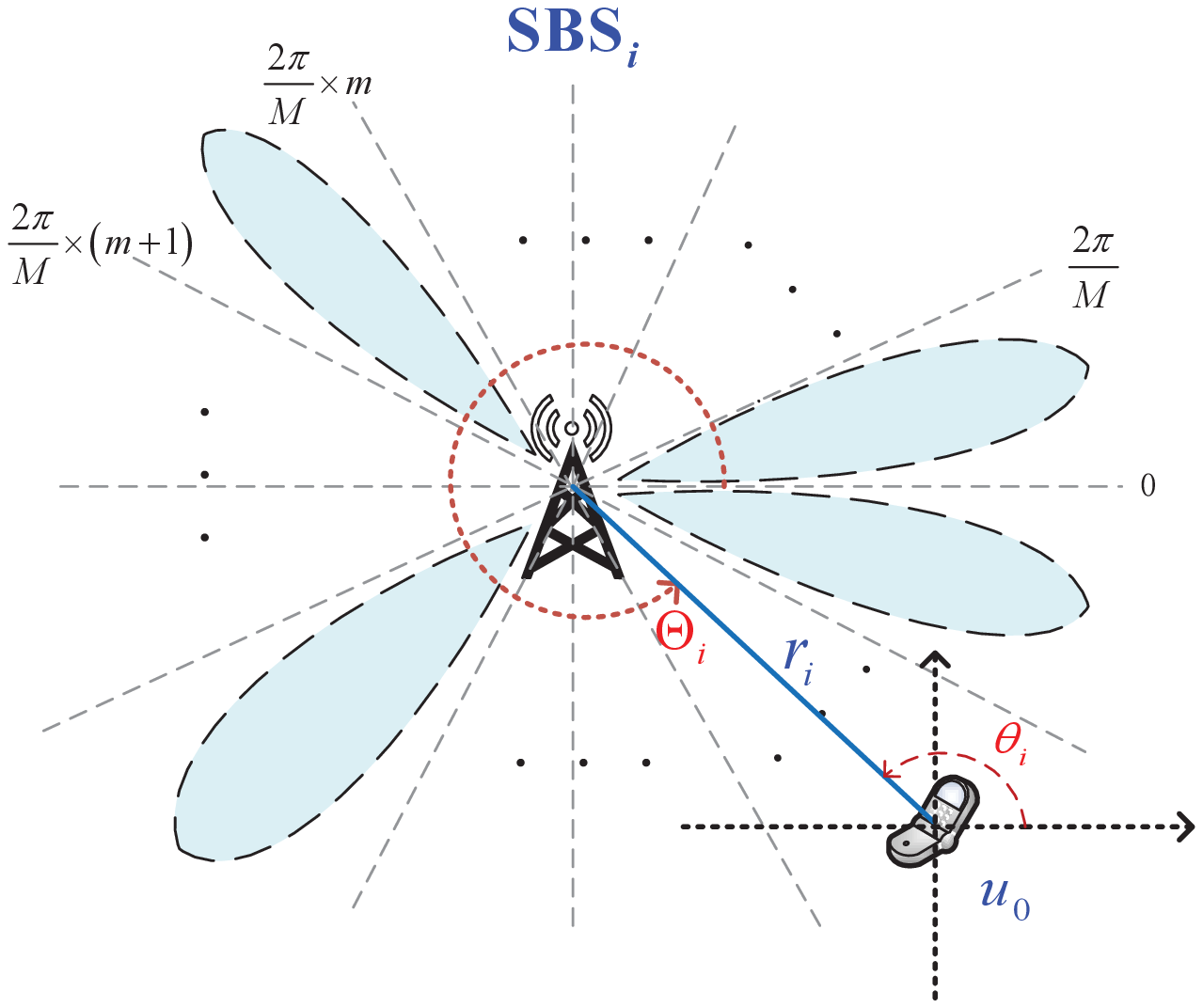}

}
\centering{}\caption{The typical anisotropic path loss scenario and model.}
\end{figure}

With the emerging mmWave technology, researchers have examined the
mmWave technology in some typical environments, \textit{e.g.}, an
urban scenario with LoS transmissions, an urban scenario with NLoS
transmissions, and the near-ground propagation in the forest. The
empirical path loss exponents for the above three scenarios were reported
as 2.1 \cite{key-29}, 3.19 \cite{key-29}, and 4 \cite{key-30},
respectively. It can be found that the path loss exponents are entirely
different in these scenarios. Considering that the three typical propagation
environments are irregularly distributed around SBSs, the links from
a user to SBSs possibly pass through different environments such that
the path loss exponents of these links are different. In this case,
a uniform path loss model cannot be used to investigate the real anisotropic
radio propagation. Thus, a multi-directional path loss model is proposed
in this paper to evaluate the radio propagation in an anisotropic
path loss scenario. A typical anisotropic path loss scenario is shown
in Fig. \ref{fig:scenario}, where a user moves from point A to point
D along the red line. The link between the user and a particular SBS
passes through the wood when the user is in the range between A and
B. The link between the user and the SBS passes through buildings
when the user is in the range between B and C. These links are NLoS.
The link between the user and the SBS is LoS when the user is in the
range between C and D.

In this model, the coverage area centered at an SBS is equally partitioned
into $M$ sections, and the path loss exponents of $M$ sections are
assumed to be different, where the value of $M$ is artificially configured
to model the anisotropic characteristic of a given propagation environment.
When $M$ is large enough, the values of the path loss exponents in
$M$ sections could be regarded as the samples of the path loss exponent
around the SBS. The sequence $\Omega_{i}=\left\{ \alpha_{im}\right\} $
denotes a stochastic process of the path loss exponent of $\textrm{SBS}_{i}$,
composed of the random variables $\alpha_{im}$ with the index $m\left(m=1,2,\ldots,M\right)$.
The anisotropic characteristic of the propagation environment could
be analyzed with statistical methods based on these $M$ samples.
Considering that the value of $M$ needs to be given to characterize
the system performance, we show in Section V by simulation that the
impact of M on the performance becomes stable when the value of M
is larger than 400.

The measured path loss exponent in real environments is mainly affected
by the obstacle distribution, the atmospheric environment, the weather
status, and diffraction and scattering effects. Since diffraction
effects are negligible and there are only a few scattering clusters
for mmWave \cite{key-25}, the impact of obstacle distribution dominates
the path loss exponent in urban environments. The obstacles in real
environments consisting of buildings, plants, cars, human bodies and
so on, are independent \cite{key-39,key-40}. Considering that the
sections in different directions are non-overlapping, the obstacles
in different sections are independent. Therefore, in this paper the
path loss exponent of the $m$-th section of $\textrm{SBS}_{i}\,(i=1,2,...)$,
denoted as $\alpha_{im}$, is assumed to be independently identically
distributed (\textit{i.i.d.}) random variables uniformly distributed
with mean $\mu$ and variance $\sigma$, and its PDF is given as
\begin{equation}
f\left(\alpha_{im}\right)=\frac{1}{2\sqrt{3}\sigma},\alpha_{im}\in\left[\mu-\sqrt{3}\sigma,\mu+\sqrt{3}\sigma\right].
\end{equation}
Note that the path loss exponent of the $m$-th section at $\textrm{SBS}_{i}$
is independent of the $m$-th section at $\textrm{SBS}_{j}\left(j\neq i\right)$,
\textit{i.e.}, $\alpha_{im}\neq\alpha_{jm}$.

A possible solution to obtain the path loss exponents of $M$ sections
of an SBS in practice is introduced as follows. In order to let the
SBS perceive the existence of the mobile phone at any time, the mobile
phone must report the channel state information (CSI) back to the
SBS regularly, or must immediately return when the SBS asks \cite{key-42}.
If the SBS is allowed to obtain the mobile phone\textquoteright s
location, the path loss exponent can be estimated by a least square\textquoteright s
method at the SBS based on the CSI and the location reported by the
mobile phone. Considering the urban scenario, every SBS serves many
users around it. The SBS collects the CSI and the locations of all
online mobile phones, and then computes the path loss exponent of
each section using the collected data from the section.

According to the 3GPP path loss models summarized in TR 38.900 \cite{key-29},
the path loss exponent in UMi-Street Canyon scenario is 2.1 when the
link distance is smaller than the breakpoint distance, and that becomes
4 when the link distance is larger than the breakpoint distance. The
characteristic mutations of path loss exponent occur at the breakpoint.
Considering a mmWave system in the small cell network that $f_{c}=24\mathrm{GHz}$,
the height of SBS is 5m, the height of user is 1.5m, the breakpoint
distance is calculated as 640m, which comfortably exceeds the average
coverage radius of an SBS given as $\frac{1}{\sqrt{\pi\lambda}}\approx56m$.
We consider the 5G fractal small cell networks with high density in
which the link distances are usually smaller than the breakpoint distance.
In this case, the difference in path loss exponents of different directions
is mainly considered in this paper. The multi-directional path loss
over the link between $\textrm{SBS}_{i}$ and the typical user $u_{0}$
is set as
\begin{equation}
L\left(x_{i}\right)=r_{i}^{-\alpha_{im}},
\end{equation}
when $\frac{2\pi}{M}\left(m-1\right)\leq\Theta_{i}<\frac{2\pi}{M}m\,\left(m=1,\ldots,M\right)$,
where $\Theta_{i}$ is the angle of the vector from $\textrm{SBS}_{i}$
to $u_{0}$, which is calculated to locate the section that the user
stays in, shown in Fig. \ref{fig:path loss model}. The angle $\Theta_{i}$
is given as
\begin{equation}
\Theta_{i}=\begin{cases}
\theta_{i}+\pi & \theta_{i}\in\left[0,\pi\right)\\
\theta_{i}-\pi & \theta_{i}\in\left[\pi,2\pi\right)
\end{cases}.
\end{equation}
Such a multi-directional path loss model can be degraded into the
isotropic path loss model with the uniform path loss exponent over
the entire plane when $\sigma=0$, \textit{i.e.}, $\alpha_{i1}=\alpha_{i2}=\cdots=\alpha_{iM}=\alpha_{c}=\mu$
as a constant with probability 1. The isotropic path loss model could
be expressed as $L^{1}\left(x_{i}\right)=r_{i}^{-\mu}$.

\begin{figure}[tbh]
\begin{centering}
\includegraphics[width=9cm]{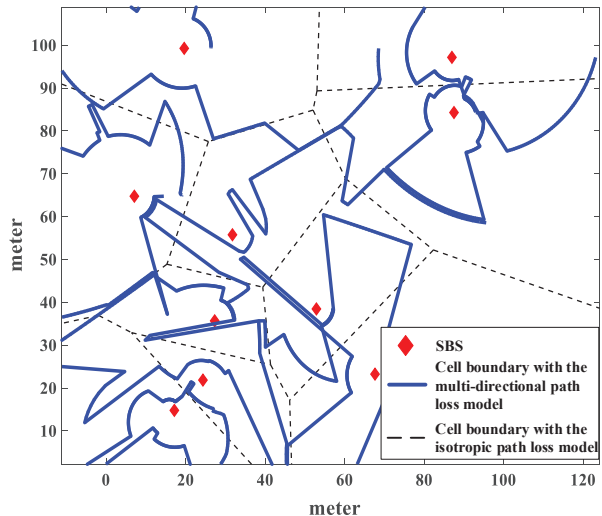}
\par\end{centering}
\caption{Illustration of 5G fractal small cell network structure with the multi-directional
path loss model ($M=5$, $\mu=4$, $\sigma=1$), compared with the
isotropic path loss model.\label{fig:PVT}}
\end{figure}

To illustrate the 5G fractal small cell network structure, assume
that a user is associated with the SBS from which the user receives
the maximum SINR, and users only suffer the path loss during propagation.
Every cell is assumed to include only one SBS. The locations at which
the user receive equal SINRs from the two adjacent SBSs, make up the
cell boundaries. For the multi-directional path loss model, the blue
solid lines in Fig. \ref{fig:PVT} depict cell boundaries inside which
a polygon corresponds a cell coverage. For the traditional isotropic
path loss model, the SBS with the maximum SINR is the closest SBS
for the user. In this case, the cell boundary can be obtained through
the Delaunay Triangulation method \cite{key-54} expressed as the
black dashed lines in Fig. \ref{fig:PVT}.

What is more, the carrier frequency factor $f_{c}$ in the path loss,
$PL=32.4+20\lg\left(r\right)+20\lg\left(f_{c}\right)\left(\textrm{dB}\right)$
\cite{key-29}, is normalized to simplify the analysis of the impact
of the anisotropic path loss, considering that the same carrier frequency
is used at all SBSs.

\subsection{Channel Model}

When the channel fading of each transmission link is assumed to be
governed by the \textit{i.i.d.} Rayleigh fading, the received signal
power at $u_{0}$ from $\textrm{SBS}_{i}$ is given as
\begin{equation}
P_{i}=P_{T}G_{i}h_{i}L\left(x_{i}\right),
\end{equation}
where $h_{i}$ denotes the power gain from Rayleigh fading between
$u_{0}$ and $\textrm{SBS}_{i}$, which is assumed to be governed
by an exponential distribution with mean 1 similar to \cite{key-7}.
For simplicity, but without loss of generality, all SBSs are assumed
to transmit with the same transmission power $P_{T}$. $G_{i}$ denotes
the antenna gain obtained from the $\textrm{SBS}_{i}$ and $u_{0}$.

For tractability of the analysis, the actual array patterns are approximated
by a sectored antenna model. Both the user and its serving SBS will
estimate the channels including angles of arrivals and fading gains,
and then adjust their antenna steering orientations accordingly to
exploit the maximum antenna gain. Errors in channel estimation are
neglected, and so are errors in time and carrier frequency synchronization
in this paper. Thus, $M_{t}$ and $M_{r}$ are the main lobe gains
of the SBS and user, respectively; $m_{t}$ and $m_{r}$ are the side
lobe gains of the SBS and user, respectively; $\phi_{t}$ and $\phi_{r}$
denote the main lobe width and side lobe width of the SBS and user,
respectively. The antenna gain of the desired link is $M_{t}M_{r}$,
and the antenna gain of the interfering link is a random variable
with the distribution shown in Table \ref{tab:antenna} \cite{key-25}.

\begin{table*}[tbh]
\caption{\label{tab:antenna}The probability distribution of antenna gain $G_{i}$}
\centering{}%
\begin{tabular}{|c|c|c|c|c|}
\hline
\textbf{Value} & $g_{1}=M_{t}M_{r}$ & $g_{2}=M_{t}m_{r}$ & $g_{3}=m_{t}M_{r}$ & $g_{4}=m_{t}m_{r}$\tabularnewline
\hline
\textbf{Probability} & $p_{1}=\frac{\phi_{t}}{2\pi}\frac{\phi_{r}}{2\pi}$ & $p_{2}=\frac{\phi_{t}}{2\pi}\left(1-\frac{\phi_{r}}{2\pi}\right)$ & $p_{3}=\left(1-\frac{\phi_{t}}{2\pi}\right)\frac{\phi_{r}}{2\pi}$ & $p_{4}=\left(1-\frac{\phi_{t}}{2\pi}\right)\left(1-\frac{\phi_{r}}{2\pi}\right)$\tabularnewline
\hline
\end{tabular}
\end{table*}

In order to obtain the best channel condition, a user is configured
to be associated with a particular SBS based on the SINR value, \textit{i.e.},
when the SINR value of the link between the user and this SBS is the
maximum. The SBS associated with the typical user $u_{0}$ is denoted
as $\textrm{SBS}_{k}$, where
\begin{equation}
k=\underset{x_{i}\in\Phi}{\mathrm{argmax}}\,\mathrm{SINR}\left(x_{i}\right),
\end{equation}
with
\begin{equation}
\mathrm{SINR}\left(x_{i}\right)=\frac{P_{i}}{I\left(x_{i}\right)+\sigma_{n}^{2}}=\frac{P_{T}M_{t}M_{r}h_{i}r_{i}^{-\alpha_{i}}}{\underset{x_{j}\in\Phi,j\neq i}{\sum}P_{T}G_{j}h_{j}r_{j}^{-\alpha_{j}}+\sigma_{n}^{2}},\label{eq:sinr}
\end{equation}
which is the SINR value of the link between $u_{0}$ and $\textrm{SBS}_{i}$
with $\sigma_{n}^{2}$ denoting the noise power and $I\left(x_{i}\right)$
denoting the aggregate interference. The path loss exponents $\alpha_{i}$
and $\alpha_{j}$ in (\ref{eq:sinr}) depend on the angles $\Theta_{i}$
and $\Theta_{j}$ of the vectors from $\textrm{SBS}_{i}$ and $\textrm{SBS}_{j}$
to the typical user, respectively.

\section{Coverage and Association Probability}

In this section, the coverage and association probabilities of 5G
fractal small cell networks are derived based on the multi-directional
path loss model. Numerical analyses of the coverage and association
probabilities are conducted to evaluate the performance impacted by
the anisotropic propagation environment.

\subsection{Coverage Probability}

We analyze the coverage performance with the multi-directional path
loss model of 5G fractal small cell networks in SINR and rate terms.
The expressions for the SINR and rate coverage probability with the
multi-directional path loss model are derived. The SINR coverage probability
to the point of view of the typical user $u_{0}$ is the probability
that the desired received SINR at $u_{0}$ is larger than a threshold
$\tau$. Considering the maximum SINR association, the SINR coverage
probability is given as
\begin{equation}
P_{C}\left(\tau\right)=\textrm{Pr}\left\{ \underset{x_{i}\in\Phi}{\max}\,\mathrm{SINR}\left(x_{i}\right)>\tau\right\} .
\end{equation}
\begin{lem}
\label{lem:1}Based on the Lemma 1 in \cite{key-31}, when the SINR
threshold $\tau>\frac{M_{t}M_{r}}{m_{t}m_{r}}$, at most one SINR
can be greater than $\tau$.
\end{lem}
\begin{IEEEproof}
See Appendix A.
\end{IEEEproof}
\begin{thm}
\label{thm:2}The SINR coverage probability for the typical user at
the origin is given as

\begin{equation}
P\!_{C}\!\left(\!\tau\!\right)\!=\!\frac{\lambda\!\text{\ensuremath{\pi}}}{\sqrt{3}\!\sigma}\!\int_{0}^{\infty}\!\!\!\int_{\mu\!-\!\sqrt{3}\sigma}^{\mu\!+\!\sqrt{3}\sigma}\!r\!\exp\!\left(\!-\!\frac{\lambda\!\pi^{2}}{\sqrt{3}\!\sigma}\!\underset{l=1}{\overset{4}{\sum}}\!p_{l}\!\int_{\mu\!-\!\sqrt{3}\sigma}^{\mu\!+\!\sqrt{3}\sigma}\!\frac{\left(\frac{\tau r^{\alpha_{i}}g_{l}}{M_{t}M_{r}}\right)^{\frac{2}{\alpha}}}{\alpha}\!\csc\!\left(\!\frac{2\pi}{\alpha}\!\right)\!\!d\!\alpha\!-\!\frac{\tau r^{\alpha_{i}}\sigma_{n}^{2}}{P_{T}M_{t}M_{r}}\!\right)\!\!d\alpha_{i}dr.\label{eq:covProg-1}
\end{equation}
\end{thm}
\begin{IEEEproof}
See Appendix B.
\end{IEEEproof}
Furthermore, we analyze the distribution of the achievable rate $R_{0}$
of 5G fractal small cell networks with the multi-directional path
loss model. The achievable rate is defined as
\begin{equation}
R_{0}=B_{W}\log_{2}\left(1+\underset{x_{i}\in\Phi}{\max}\,\mathrm{SINR}\left(x_{i}\right)\right),
\end{equation}
where $B_{W}$ is the bandwidth assigned to the typical user $u_{0}$.
The rate coverage probability is the probability that the achievable
rate of the typical user is larger than the threshold $\gamma$, which
is expressed as
\begin{align}
P_{R}\left(\gamma\right) & =\Pr\left\{ R_{0}>\gamma\right\} =\Pr\left\{ B_{W}\log_{2}\left(1+\underset{x_{i}\in\Phi}{\max}\,\mathrm{SINR}\left(x_{i}\right)\right)>\gamma\right\} \nonumber \\
 & =\Pr\left\{ \underset{x_{i}\in\Phi}{\max}\,\mathrm{SINR}\left(x_{i}\right)>2^{\frac{\gamma}{B_{W}}}-1\right\} =P_{C}\left(2^{\frac{\gamma}{B_{W}}}-1\right).
\end{align}

To express the impact of the multi-directional path loss model on
the coverage performance of the 5G fractal small cell network, numerical
results of the SINR and rate coverage probabilities are presented
considering the factors including the SINR threshold $\tau$, the
rate threshold $\gamma$, the SBS density $\lambda$, and the mean
$\mu$ and the variance $\sigma$ of the path loss exponent. The default
parameters are set as: $P_{T}=30\textrm{dBm}$, $B_{W}=500\mathrm{MHz}$,
$\sigma_{n}^{2}=-174\unitfrac{\textrm{dBm}}{Hz}\times500\textrm{MHz}=-87\textrm{dBm}$
\cite{key-41}, $M_{t}=M_{r}=10\textrm{dB}$, $m_{t}=m_{r}=0\textrm{dB}$,
$\phi_{t}=7\textdegree$, $\phi_{r}=60\textdegree$ \cite{key-3},
$\lambda=10^{-4}$, $\tau=20\textrm{dB}$ and $\mu=4$ in this paper
unless noted otherwise. The large variance $\sigma$ implies the significant
difference among the path loss exponents of SBSs in the 5G fractal
small cell network.

\begin{figure}[tbh]
\begin{centering}
\includegraphics[width=9cm]{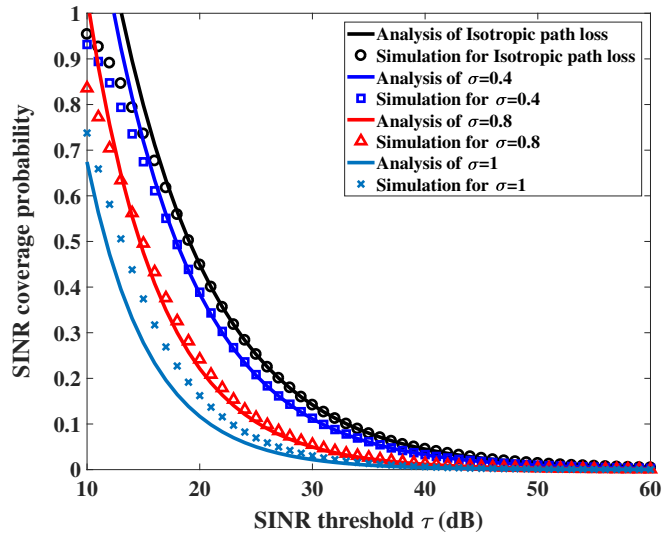}
\par\end{centering}
\caption{\label{fig:The-coverage-probability-tau}The SINR coverage probability
with respect to the SINR threshold $\text{\ensuremath{\tau}}$, considering
different variances of the path loss exponent.}
\end{figure}

Fig. \ref{fig:The-coverage-probability-tau} illustrates the SINR
coverage probability with respect to the SINR threshold $\tau$, considering
different variances $\sigma$ of the path loss exponent. The SINR
coverage probability decreases with the increase of the SINR threshold.
When the variance is equal to zero, \textit{i.e.}, $\sigma=0$, the
path loss exponent is a constant with probability 1, and is equal
to the mean $\mu$. Thus, the case with $\sigma=0$ represents the
isotropic path loss model in this paper. When the SINR threshold is
fixed, the SINR coverage probability decreases with the increase of
the variance $\sigma$, which indicates that the anisotropic path
loss model has a negative influence on the coverage performance of
the small cell network. According to the proof of Lemma \ref{lem:1},
the result shows numerically that this SINR coverage probability model
holds very accurately.

\begin{figure}[tbh]
\begin{centering}
\includegraphics[width=9cm]{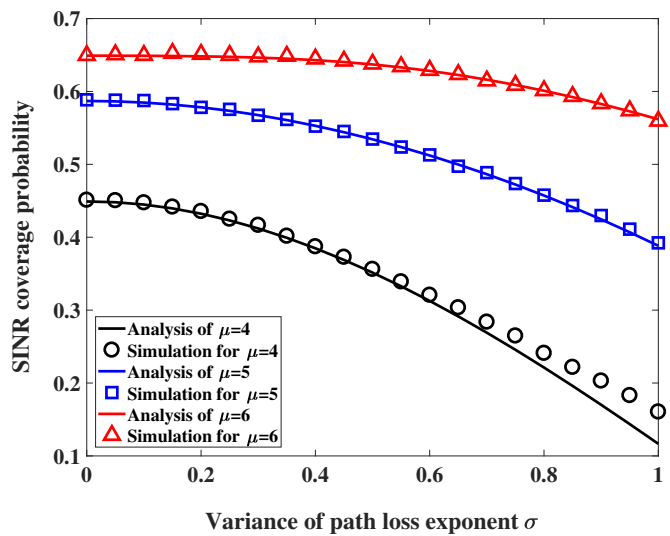}
\par\end{centering}
\caption{\label{fig:The-coverage-probability-sigma}The SINR coverage probability
with respect to the variance of the path loss exponent $\sigma$,
considering different mean values $\mu$.}
\end{figure}

Fig. \ref{fig:The-coverage-probability-sigma} shows the SINR coverage
probability with respect to the variance of the path loss exponent
$\sigma$, considering different mean values $\mu$. The larger mean
value $\mu$ results in that the SINR coverage probability increases,
which implies that the coverage performance is better in the higher
path loss attenuation environment due to the more severe attenuation
of the interference signal. When the mean $\mu$ is fixed, the SINR
coverage probability decreases with the increase of the variance $\sigma$,
which shows that the coverage performance is better in the more stable
propagation environment, especially in the isotropic propagation environment
with the case $\sigma=0$. It is because the path loss exponents of
a part of interference links are so small in the anisotropic propagation
environment that increase the aggregate interference power compared
to the isotropic path loss model.

\begin{figure}[tbh]
\begin{centering}
\includegraphics[width=9cm]{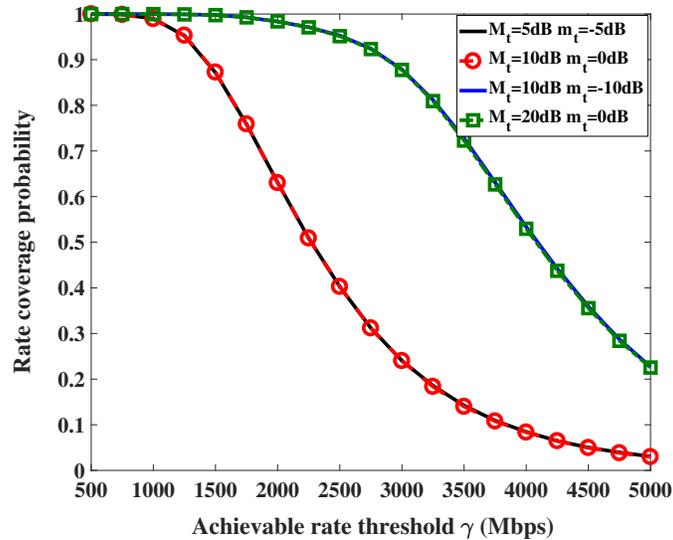}
\par\end{centering}
\begin{centering}
\caption{\label{fig:rate-coverage}The rate coverage probability with respect
to the achievable rate threshold $\gamma$, considering different
antenna gains.}
\par\end{centering}
\end{figure}

Fig. \ref{fig:rate-coverage} illustrates the rate coverage probability
with respect to the achievable rate threshold $\gamma$, considering
different antenna gains. The rate coverage probability increases with
the larger ratio of the main lobe gain to the side lobe gain $\frac{M_{t}}{m_{t}}$.
It can be found in the figure that the red and black lines with $\frac{M_{t}}{m_{t}}=10\mathrm{dB}-0\mathrm{dB}=5\mathrm{dB}-(-5)\mathrm{dB}=10\mathrm{dB}$
ate below the green and blue lines with $\frac{M_{t}}{m_{t}}=20\mathrm{dB}-0\mathrm{dB}=10\mathrm{dB}-(-10)\mathrm{dB}=20\mathrm{dB}$.
What\textquoteright s more, the rate coverage probability is independent
from the values of and when the ratio of the main lobe gain to the
side lobe gain $\frac{M_{t}}{m_{t}}$ is fixed.

\subsection{Association Probability}

Some existing analysis for the handoff performance works based on
the assumption that the nearest neighbor SBS to be the one with the
highest SINR. When the isotropic path loss model is adopted, it can
be proved that a user is covered by the nearest SBS with the maximum
probability. However, the association probability with an LoS SBS
may be greater than that with an NLoS SBS even when the distance between
the LoS SBS and the user is larger than the distance between the NLoS
SBS and the user in a complex propagation environment, especially
in small cell networks \cite{key-33}. In this section, association
probabilities for the typical user with the different SBSs in 5G fractal
small cell networks are analyzed based on the multi-directional path
loss model to address the above situations.

In this paper, a user is assumed to be associated with the SBS that
leads to the maximum received SINR. The association probability with
respect to the distance $r$ between $u_{0}$ and the desired SBS
$\textrm{SBS}^{r}$ is expressed as
\begin{equation}
P_{A}\left(r\right)=\Pr\left\{ \mathrm{SINR}\left(r\right)=\underset{x_{i}\in\Phi}{\max}\,\mathrm{SINR}\left(x_{i}\right)\right\} \text{,}
\end{equation}
\begin{equation}
\mathrm{SINR}\left(r\right)=\frac{P_{T}M_{t}M_{r}h_{r}r^{-\alpha_{r}}}{I\left(r\right)+\sigma_{n}^{2}},
\end{equation}
where $h_{r}$, and $\alpha_{r}$ denote the antenna gain, power gain
from Rayleigh fading, and path loss exponent of the link between $u_{0}$
and the desired SBS $\textrm{SBS}^{r}$, respectively. $I\left(r\right)$
is the received interference of the user $u_{0}$.

Considering the difficulty of calculating the association probability
due to the complex expression of SINR, an approximation that greatly
simplifies the analysis is that the user is associated with the SBS
that leads to the maximum received power. The following Lemma shows
that the approximation has the high accuracy, and the accuracy increases
with the narrower main lobes at the transmitter and receiver.
\begin{lem}
The probability that $\mathrm{SINR}\left(x_{k}\right)$ is the maximum
is at least $\left(1-\frac{\phi_{t}}{360}\right)\left(1-\frac{\phi_{r}}{360}\right)$
when the corresponding received signal power $P_{k}$ is maximum.
\end{lem}
\begin{IEEEproof}
See Appendix C.
\end{IEEEproof}
Note that the system level simulations are given for the probability
in Lemma 3, expressed as $\Pr\left\{ \mathrm{SINR}\left(r\right)|P\left(r\right)\right\} $.
The results are shown in Table II and imply that the probability is
higher when the narrower main lobes are equipped at the transmitter
and receiver. In this case, the probability that $\mathrm{SINR}\left(r\right)$
is the maximum can be approximated by the probability that $P\left(r\right)$
is the maximum. Based on Lemma 3, the association probability is further
expressed as
\begin{equation}
P_{A}\left(r\right)\approx\Pr\left\{ P\left(r\right)=\underset{x_{i}\in\Phi}{\max}\,P_{i}\right\} ,
\end{equation}
where $P\left(r\right)$ denotes the received signal power at $u_{0}$
from $\textrm{SBS}^{r}$, expressed as $P\left(r\right)=G_{r}h_{r}r^{-\alpha_{r}}$.

\begin{table}[tbh]
\caption{The probability that $\mathrm{SINR}\left(r\right)$ is the maximum
when $P\left(r\right)$ is the maximum $\left(r=50\mathrm{m}\right)$.}
\centering{}%
\begin{tabular}{|c|c|c|c|c|c|c|}
\hline
\textbf{$\phi_{t}$} & $7\textdegree$ & $20\textdegree$ & $60\textdegree$ & $60\textdegree$ & $20\textdegree$ & $7\textdegree$\tabularnewline
\hline
\textbf{$\phi_{r}$} & $60\textdegree$ & $60\textdegree$ & $60\textdegree$ & $20\textdegree$ & $10\textdegree$ & $10\textdegree$\tabularnewline
\hline
\hline
\textbf{$\Pr\left\{ \mathrm{SINR}\left(r\right)|P\left(r\right)\right\} $} & 0.9176 & 0.9035 & 0.8822 & 0.9036 & 0.9553 & 0.9709\tabularnewline
\hline
\end{tabular}
\end{table}
\begin{thm}
The association probability with respect to the distance $r$ between
$u_{0}$ and the desired SBS $\textrm{SBS}^{r}$ is approximately
given as
\end{thm}
\begin{equation}
P_{A}\!\left(\!r\!\right)\!=\!\underset{m=1}{\overset{4}{\sum}}\!\frac{p_{m}}{2\sqrt{3}\sigma}\!\int_{0}^{\infty}\!\int_{\mu\!-\!\sqrt{3}\sigma}^{\mu\!+\!\sqrt{3}\sigma}\!\exp\!\left(\!-\!h_{r}\!-\!\frac{\lambda\pi}{\sqrt{3}\sigma}\!\underset{n=1}{\overset{4}{\sum}}\!p_{n}\!\int_{\mu\!-\!\sqrt{3}\sigma}^{\mu\!+\!\sqrt{3}\sigma}\!\frac{\left(\frac{g_{m}}{g_{n}}h_{r}r^{-\alpha_{r}}\right)^{-\!\frac{2}{\alpha_{i}}}\!\varGamma\left(\frac{2}{\alpha_{i}}\right)}{\alpha_{i}}\!d\alpha_{i}\!\right)\!d\alpha_{r}dh_{r}.\label{eq:assPro}
\end{equation}
\begin{IEEEproof}
See Appendix\textbf{ }D.
\end{IEEEproof}
Fig. \ref{fig:assPro_r_sigma} shows the association probability with
respect to the distance $r$ between $u_{0}$ and the desired SBS
$\textrm{SBS}^{r}$, considering different variances $\sigma$. When
the variance of the path loss exponent is fixed, the association probability
decreases with the increase of the distance $r$ between $u_{0}$
and the desired SBS $\textrm{SBS}^{r}$. When the distance $r$ is
smaller than 80 meters, the association probability increases with
the decrease of the variance $\sigma$. When the distance $r$ is
larger than 160 meters, the association probability increases with
the increase of the variance $\sigma$. The realistic association
of users in small cell networks is more complicated than the NBA scheme.
The result also implies that the nearest base station association
scheme is not the desired method to obtain the best channel gain in
small cell networks with mmWave. Thus, we need to pay more attention
to the novel association scheme when analyzing the performance of
small cell networks with mmWave. Fig. \ref{fig:assPro_r_lambda} shows
the association probability with respect to the distance $r$ between
$u_{0}$ and the desired SBS $\textrm{SBS}^{r}$, considering different
intensities of SBSs $\lambda$. The variance of the path loss exponent
is set as $\sigma=0.8$. When the distance $r$ is fixed, the association
probability decreases with the increase of the intensity of SBSs.
What's more, we now focus on the approximation in Lemma 3 by comparing
the theoretical and simulated results for association probability
in Fig. \ref{fig:assPro_sigma}. As expected, the simulated and analytical
results match reasonably well.

\begin{figure}[tbh]
\begin{centering}
\includegraphics[width=9cm]{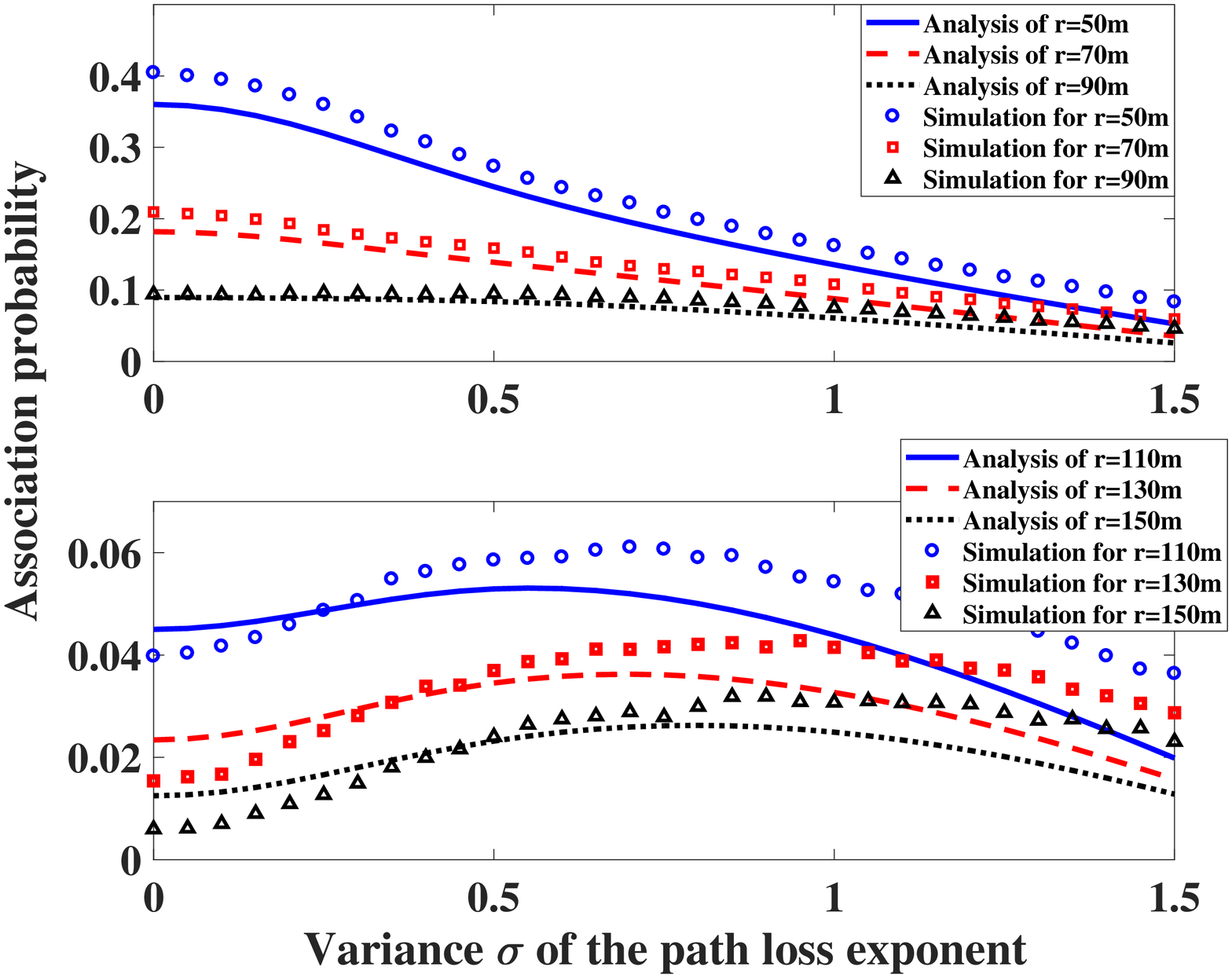}
\par\end{centering}
\caption{\label{fig:assPro_r}The association probability with respect to the
variances $\sigma$, considering different distances $r$ between
$u_{0}$ and $\textrm{SBS}^{r}$.}
\end{figure}

Fig. \ref{fig:assPro_r} shows the association probability with respect
to the variances $\sigma$, considering different distances $r$ between
$u_{0}$ and $\textrm{SBS}^{r}$. The mean value of the path loss
exponent in this figure is equal to 5 to ensure that the minimum value
of the path loss exponent is larger than 2. When the distance $r$
between $u_{0}$ and the desired SBS $\textrm{SBS}^{r}$ is shorter
than 90m, the association probability decreased with the increase
of the variance $\sigma$ of the path loss exponent. When the distance
is larger than 110m, the association probability has a peak value
at particular variances.

\begin{figure*}[tbh]
\begin{centering}
\subfloat[\label{fig:assPro_r_sigma}]{\includegraphics[width=7cm]{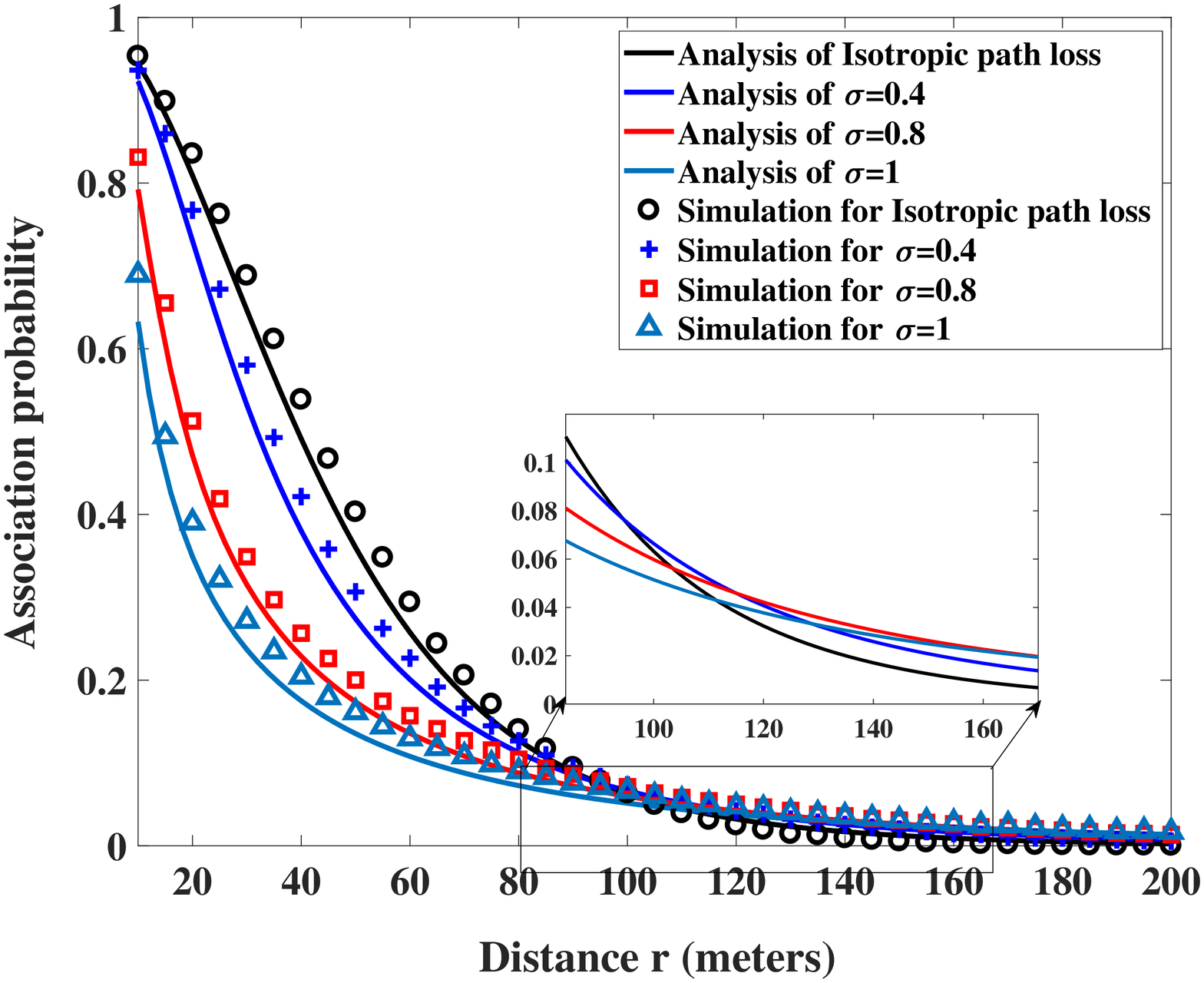}

}\subfloat[\label{fig:assPro_r_lambda}]{\includegraphics[width=7cm]{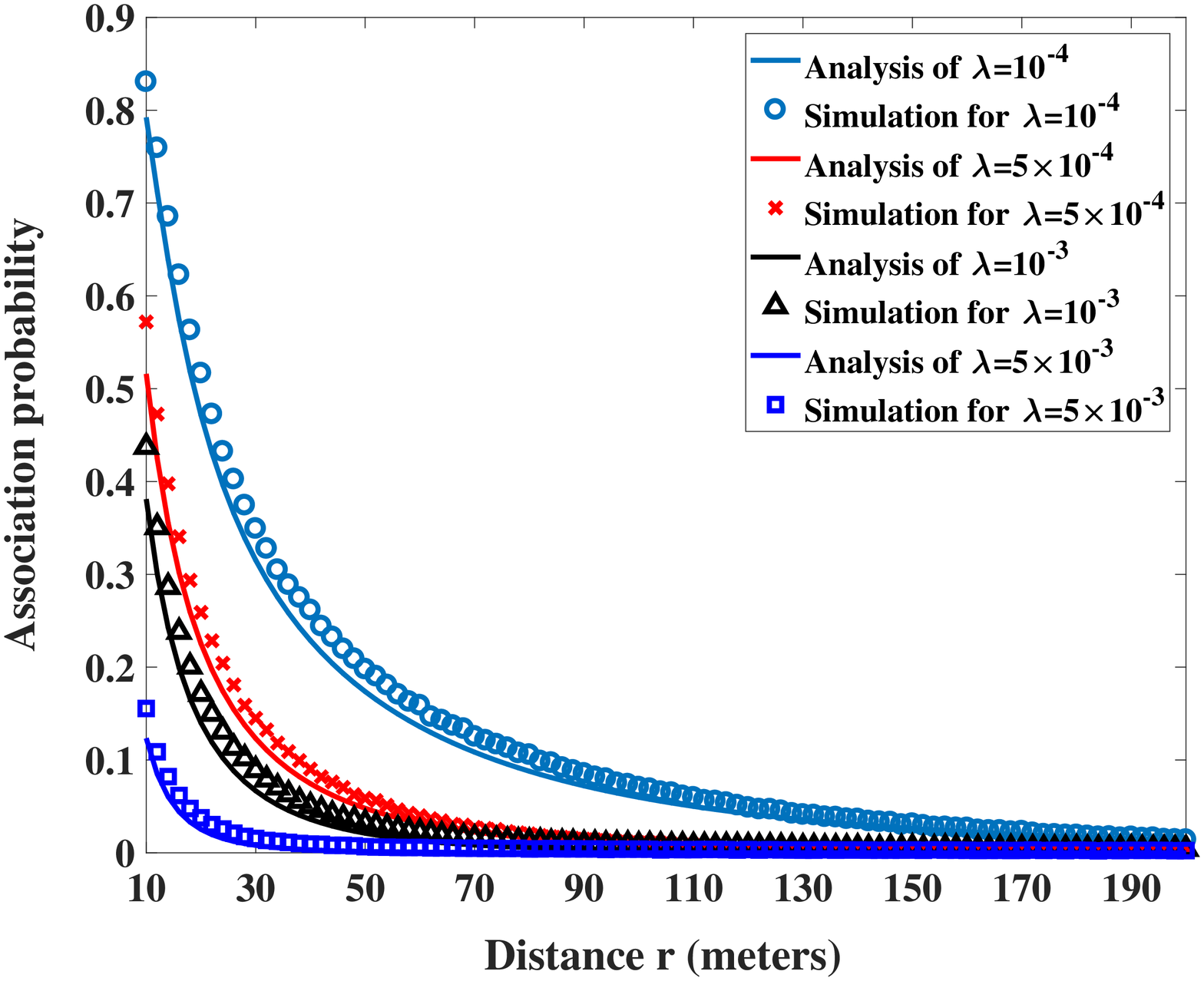}

}
\par\end{centering}
\caption{\label{fig:assPro_sigma}The association probability with respect
to the distance $r$ between $u_{0}$ and the desired SBS $\textrm{SBS}^{r}$,
considering different variances of the path loss exponent and the
intensities of SBSs.}
\end{figure*}

\section{Handoff Performance}

In particular, the handoff is crucial for keeping users connected
while moving around. Frequent handoff in small cell networks affects
not only service quality of users but also network performance, such
as throughput and energy efficiency. Conditional handoff analysis
is based on the isotropic propagation environment. However, the negative
influence of the anisotropic environment on the handoff performance
is predictable in future small cell networks. In this section, the
handoff probability and handoff rate are derived based on the user
random walk mobility model to investigate the impact of the multi-directional
path loss model.

\subsection{Handoff Probability}

Assumed that the serving SBS of the typical user $u_{0}$ at the location
$l_{0}$ is denoted as $\mathrm{\textrm{SBS}}^{r}$, \textit{i.e.},
the SINR of $\mathrm{\textrm{SBS}}^{r}$ at the location $l_{0}$
is the maximum. Let $l_{0}$ and $l_{1}$ respectively denote the
locations at the start and end of a short time movement, and the moving
time is the time interval between the end of the last handoff decision
and the begin of the next one. 3GPP Standard defines six handoff events
for cellular networks \cite{key-34}. In this paper, the analysis
of handoff performance with the multi-directional path loss model
focuses on the most common Event A2. The Event A2 occurs when the
SINR of the serving SBS becomes smaller than a threshold $\tau_{h}$,
and the trigger condition can be expressed as
\begin{equation}
\mathrm{SINR}\left(\tilde{r}\right)<\tau_{h}-\tau_{\textrm{hys}},\label{eq:handoff-con1}
\end{equation}
where $\tau_{\textrm{hys}}$ is a hysteresis parameter added for reducing
redundant handoffs (e.g. ping-pong effect), and $\tilde{r}$ is the
distance between the serving SBS and the user at the new location
$l_{1}$. Once inequality (\ref{eq:handoff-con1}) is satisfied for
the user, an Event A2 handoff is triggered, and the user needs to
select a suitable target SBS. In this case, the handoff probability,
$P_{H}\left(vt\right)$, denotes the probability that a handoff occurs
when the user moves a distance $vt$, which is expressed as $P_{H}\left(vt\right)=\int_{0}^{\infty}\Pr\left\{ \textrm{A2}\right\} dr$.
When the distance between the serving SBS and the user at location
$l_{0}$ is $r$, the probability that the Event A2 occurs is given
as
\begin{equation}
\Pr\left\{ \textrm{A2}\right\} =\Pr\left\{ \mathrm{SINR}\left(\tilde{r}\right)<\tau_{h}-\tau_{\textrm{hys}}\right\} .
\end{equation}

\begin{figure*}[tbh]
\subfloat[\label{fig:mobility} The random walk mobility model.]{\begin{raggedright}
\includegraphics[width=9cm]{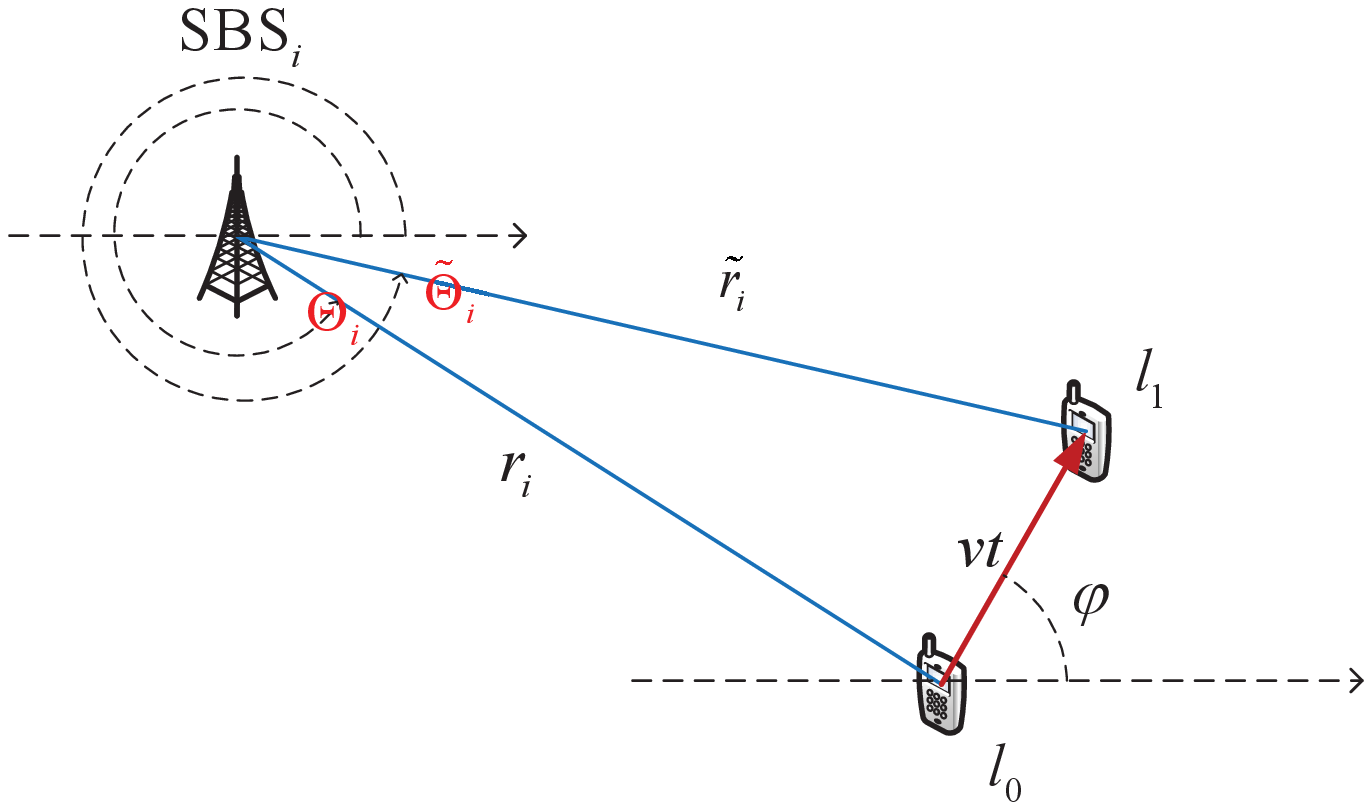}
\par\end{raggedright}
} \subfloat[\label{fig:case2}The scenario that user moves at the direction $\varphi=\Theta^{r}+\frac{\pi}{2}$.]{\begin{raggedleft}
\includegraphics[width=6cm]{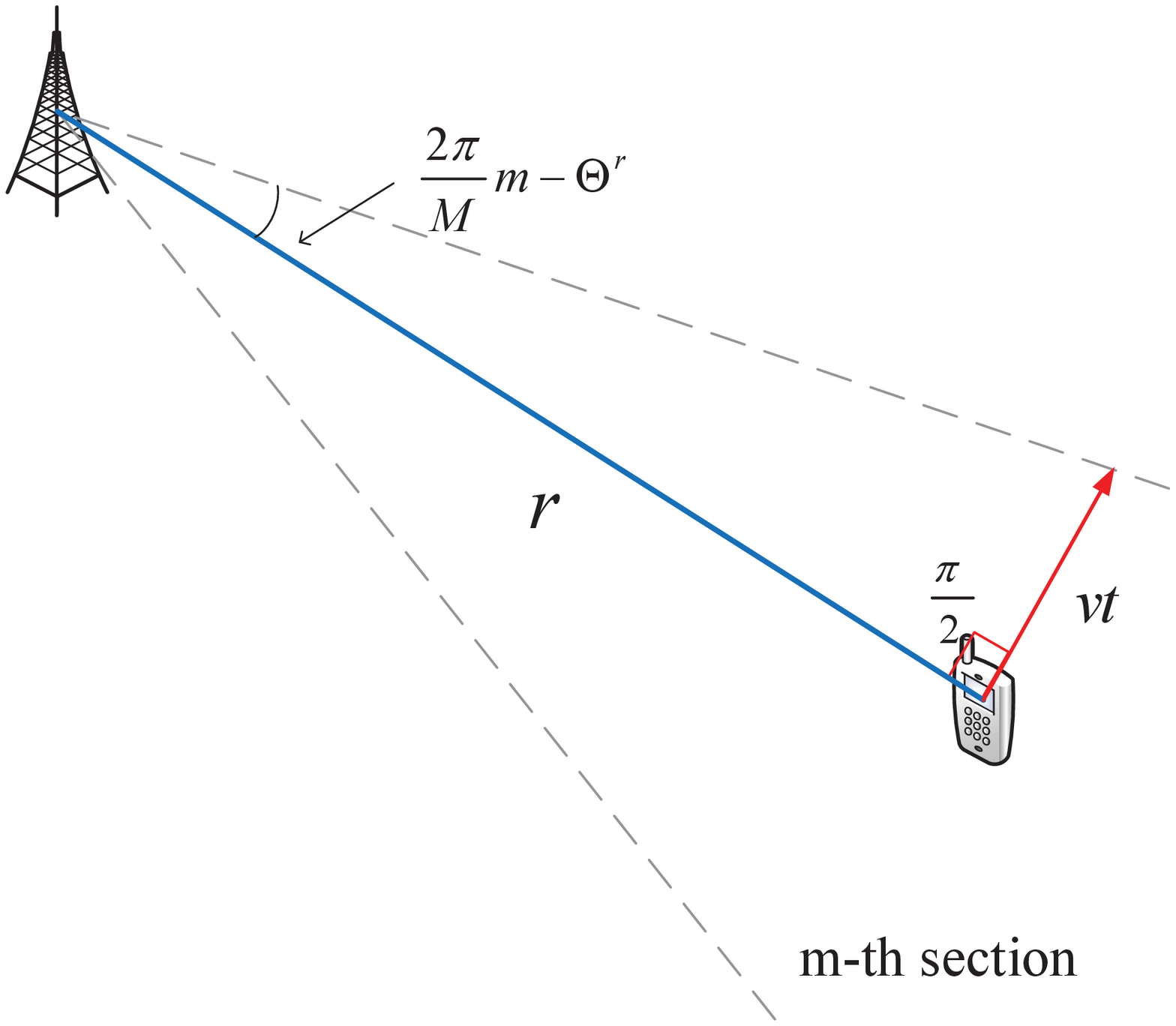}
\par\end{raggedleft}
}

\caption{The random walk mobility model with the multi-directional path loss
model.}
\end{figure*}

Before calculating the handoff probability, the relationship between
the locations $l_{0}$ and $l_{1}$ are considered. The movement of
a user is assumed to be governed by a random walk mobility model in
this paper. The user moves from its original location to a new location
by randomly choosing a direction $\varphi$ and a speed $v$, where
the speed and direction are both chosen from pre-defined ranges $\left[v_{min},v_{max}\right]$
and $\left[0,2\pi\right)$, respectively. Such a random walk mobility
model is shown in Fig. \ref{fig:mobility}, where $u_{0}$ moves from
the location $l_{0}$ to $l_{1}$, $\Theta_{i}$ is the angle between
the horizontal line and the line crossing through $l_{0}$ and $\textrm{SBS}_{i}$,
$\tilde{\Theta}_{i}$ is the angle between the horizontal line and
the line crossing through $l_{1}$ and $\textrm{SBS}_{i}$, $r_{i}$
and $\tilde{r}_{i}$ are the distances between $\textrm{SBS}_{i}$
and $l_{0}$, $l_{1}$, respectively, $t$ is the user moving time,
and $vt$ is the distance of the user moving with the speed $v$ and
time $t$. The relationship between the locations $l_{0}$ and $l_{1}$
is expressed as
\begin{equation}
\tilde{r}_{i}^{2}=r_{i}^{2}+\left(vt\right)^{2}+2vtr_{i}\cos\left(\varphi-\Theta_{i}\right),
\end{equation}
\begin{equation}
\tilde{\Theta}_{i}=\Theta_{i}+\arctan\left(\frac{vt\sin\left(\varphi-\Theta_{i}\right)}{r+vt\cos\left(\varphi-\Theta_{i}\right)}\right).
\end{equation}

Two special cases are considered in the derivation. One case is that
the user moves away from the serving SBS: $\textrm{SBS}^{r}$, \textit{i.e.},
$\varphi=\Theta^{r}$, where $\Theta^{r}$ is the angle between the
horizontal line and the line crossing through $l_{0}$ and $\textrm{SBS}^{r}$.
The other case is that the user moves to the direction perpendicular
to the line crossing through $l_{0}$ and $\textrm{SBS}^{r}$, \textit{i.e.},
$\varphi=\Theta^{r}+\frac{\pi}{2}$.

\textit{Case 1:} $\varphi=\Theta^{r}$. In this case, the path loss
exponent of the link between the user and the serving SBS is unchanged.
When the user moves a distance $vt$, the received SINR of the serving
SBS at the user is given as
\begin{equation}
\mathrm{SINR}^{0}\left(\tilde{r}\right)=\frac{P_{T}M_{t}M_{r}\tilde{h}_{r}\left(r+vt\right)^{-\alpha_{r}}}{I\left(r+vt\right)+\sigma_{n}^{2}},
\end{equation}
where $\tilde{h}_{r}$ is the power gain from Rayleigh fading at the
new location, which is independent of $h_{r}$.

The handoff probability is expressed as
\begin{equation}
P_{H}^{0}\left(vt\right)\!=\!\int_{r}\!\mathbb{E}_{\alpha_{r}}\!\left[\!\Pr\!\left\{ \!\mathrm{SINR}^{0}\left(\tilde{r}\right)\!<\!\tau_{h}\!-\!\tau_{\textrm{hys}}\!|\!\alpha_{r}\right\} P_{A}\left(r\!|\!\alpha_{r}\right)\right]dr,\label{eq:hpro1}
\end{equation}
with the conditional association probability following (\ref{eq:assPro})
\begin{align}
P_{A}\left(r|\alpha_{r}\right) & =\underset{m=1}{\overset{4}{\sum}}p_{m}\int_{0}^{\infty}\exp\left(-h_{r}-\frac{\lambda\pi}{\sqrt{3}\sigma}\underset{n=1}{\overset{4}{\sum}}p_{n}\times\int_{\mu-\sqrt{3}\sigma}^{\mu+\sqrt{3}\sigma}\frac{\left(\frac{g_{m}}{g_{n}}h_{r}r^{-\alpha_{r}}\right)^{-\frac{2}{\alpha_{i}}}\varGamma\left(\frac{2}{\alpha_{i}}\right)}{\alpha_{i}}d\alpha_{i}\right)dh_{r}.\label{eq:pa_r}
\end{align}
\begin{thm}
\label{thm:5}When the user moves to the direction $\varphi=\Theta^{r}$,
the handoff probability is
\begin{align}
 & P_{H}^{0}\left(vt\right)=\!\int_{0}^{\infty}\!\int_{\mu\!-\!\sqrt{3}\sigma}^{\mu\!+\!\sqrt{3}\sigma}\!\left(\!1\!-\!\exp\left(\!-\!\left(\frac{\tau_{h}\!-\!\tau_{\textrm{hys}}}{P_{T}M_{t}M_{r}}\right)\sigma_{n}^{2}\left(r\!+\!vt\right)^{\alpha_{r}}\right)\right.\nonumber \\
 & \left.\times\!\exp\left(\!-\frac{\lambda\pi^{2}}{\sqrt{3}\sigma}\!\underset{l=1}{\overset{4}{\sum}}p_{l}\!\int_{\alpha}\!\frac{\left(\frac{\left(\tau_{h}\!-\!\tau_{\textrm{hys}}\right)\left(r\!+\!vt\right)^{\alpha_{r}}g_{l}}{M_{t}M_{r}}\right)^{\frac{2}{\alpha}}\!\csc\left(\frac{2\pi}{\alpha}\right)}{\alpha}\!d\alpha\right)\right)\label{eq:handoff1}\\
 & \times\!\left(\!\underset{m\!=\!1}{\overset{4}{\sum}}\!p_{m}\!\int_{0}^{\infty}\!\exp\!\left(\!-\!h_{r}\!-\!\frac{\lambda\pi}{\sqrt{3}\sigma}\!\underset{n=1}{\overset{4}{\sum}}\!p_{n}\!\times\!\int_{\mu\!-\!\sqrt{3}\sigma}^{\mu\!+\!\sqrt{3}\sigma}\!\frac{\left(\frac{g_{m}}{g_{n}}h_{r}r^{-\alpha_{r}}\right)^{-\frac{2}{\alpha_{i}}}\varGamma\left(\frac{2}{\alpha_{i}}\right)}{\alpha_{i}}\!d\alpha_{i}\!\right)\!dh_{r}\!\right)\!d\alpha_{r}dr.\nonumber
\end{align}
\end{thm}
\begin{IEEEproof}
See Appendix\textbf{ }E.
\end{IEEEproof}
\textit{Case 2:} $\varphi=\Theta^{r}+\frac{\pi}{2}$. As shown in
Fig. \ref{fig:case2}, the path loss exponent is unchanged when the
moving distance satisfies $vt\leq r\tan\left(\frac{2\pi}{M}m-\Theta^{r}\right)$.
When the moving distance is larger than $r\tan\left(\frac{2\pi}{M}m-\Theta^{r}\right)$,
the path loss exponent changes. Then, the received SINR of the serving
SBS at the user is given as
\begin{equation}
\mathrm{SINR}^{90}\left(\tilde{r}\right)=\frac{P_{T}M_{t}M_{r}\tilde{h}_{r}\left(r^{2}+\left(vt\right)^{2}\right)^{-\frac{\tilde{\alpha}_{r}}{2}}}{I\left(\sqrt{r^{2}+\left(vt\right)^{2}}\right)+\sigma_{n}^{2}},
\end{equation}
where $\tilde{\alpha}_{r}$ denotes the new path loss exponent of
the link between the serving SBS and the user located at sections
except for the $m$-th section. Given that the distance $r$ between
the user and the serving SBS is fixed and known, angle $\Theta^{r}$
is assumed to be uniformly distributed in $\left[0,2\pi\right)$ and
independent of the distance $r$ \cite{key-35}. In this case, $\varDelta\Theta=\frac{2\pi}{M}m-\Theta^{r}$
is uniformly distributed in $\left[0,\frac{2\pi}{M}\right)$. The
handoff probability is expressed as
\begin{align}
 & P_{H}^{90}\left(vt\right)\!=\!\frac{M}{2\pi}\!\int_{0}^{\frac{2\pi}{M}}\!\left(\!\int_{0}^{\frac{vt}{\tan\left(\varDelta\Theta\right)}}\!\Pr\left\{ \!\mathrm{SINR}^{90}\left(\tilde{r}\right)\!<\!\tau_{h}\!-\!\tau_{\textrm{hys}}\!\right\} \!P_{A}\!\left(r\right)\!dr\right.\nonumber \\
 & \left.+\int_{\frac{vt}{\tan\left(\varDelta\Theta\right)}}^{\infty}\mathbb{E}_{\alpha_{r}}\left[\Pr\left\{ \mathrm{SINR}^{90}\left(\tilde{r}\right)<\tau_{h}-\tau_{\textrm{hys}}|\tilde{\alpha}_{r}=\alpha_{r}\right\} \times P_{A}\left(r|\alpha_{r}\right)\right]\right)drd\varDelta\Theta.
\end{align}
The detailed calculation is similar to the Theorem \ref{thm:5}, which
is omitted here. The final expression of the handoff probability with
$\varphi=\Theta^{r}+\frac{\pi}{2}$ is not shown here.

\subsection{Handoff Rate}

To evaluate the effect of the anisotropic path loss on the handoff
performance, the handoff overhead should be taken into consideration
for 5G fractal small cell networks. During handoff execution, a user
releases the serving SBS and establishes a new association with the
target SBS through control protocols with the two related SBSs. The
handoff overhead is contributed by control signals transmitted between
the MS and the SBSs. Hence, the overall handoff overhead is directly
proportional to the handoff rates.

To easily evaluate the handoff overhead, the handoff rate is defined
as the average number of handoffs per unit time. According to the
handoff control protocols \cite{key-36}, the handoff decision procedures
are periodically executed, where the signal detection and information
exchanging between the user and the original associated SBS are conducted
in each given period. Such a period is denoted as the detection interval
$t_{d}$. Thus, in this paper Monte-Carlo simulations are utilized
to analyze the handoff rate of 5G fractal small cell networks, with
the multi-directional path loss model.

We consider a small cell network deployed in an urban area, and the
network consists of a user and a varying number of SBSs. The user
is initially located at the central of a circular area with radius
equal to 1000m, and SBSs are randomly distributed in the area. Some
parameters related to system model are the same as those in Section
III. The user moves to a random direction at a speed $v$. In each
period $t_{d}$, the channel state information is reported to the
serving SBS by the user, and a handoff decision mechanism is executed.
We assume that a perfect initial cell search can be performed, and
thus the user can discover SBSs correctly when a handoff occurs. The
number of handoffs in 1000 seconds is recorded to calculate the handoff
rate.

\section{SIMULATION RESULTS}

The handoff probability and handoff rate are analyzed for 5G fractal
small cell networks with the multi-directional path loss model (where
the uniform path loss model is treated as a special case with $\sigma=0$)
in this section. The default parameters are configured as in Section
III. The SINR threshold for handoff with a hysteresis parameter is
configured as $\tau_{h}-\tau_{\textrm{hys}}=0\mathrm{\textrm{dB}}$.

\begin{figure}[tbh]
\begin{centering}
\includegraphics[width=9cm]{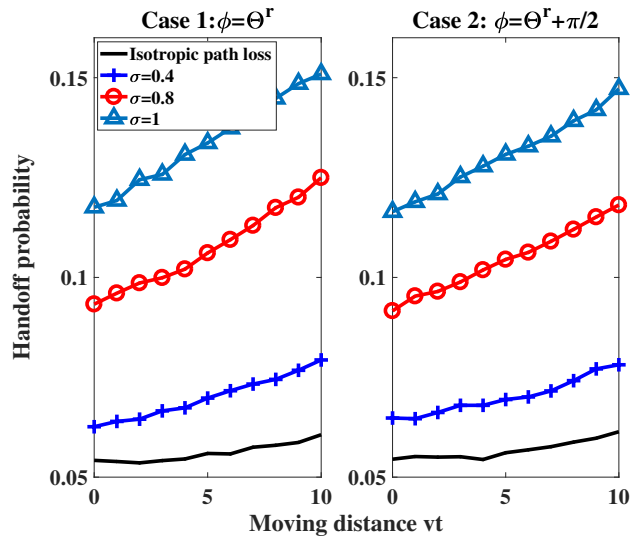}
\par\end{centering}
\caption{\label{fig:handPro_vt_sigma}The handoff probability with respect
to the moving distance $vt$ in the two special cases, considering
different variances of the path loss exponent.}
\end{figure}

Fig. \ref{fig:handPro_vt_sigma} shows the handoff probability with
respect to the moving distance $vt$ in the two special cases, considering
different variances of the path loss exponent. In this figure, the
number of sections of each SBS is configured as $M=3$. The handoff
probability increases with the increase of the moving distance in
a detection interval. And the handoff probability increases with the
increase of the variance of the path loss exponent. The difference
between the handoff probabilities in the case 1 and case 2 could be
ignored at the same condition with moving distance and variance of
the path loss exponent. The results imply that the direction of the
user movement has no influence on the handoff performance.

\begin{figure}[tbh]
\begin{centering}
\includegraphics[width=9cm]{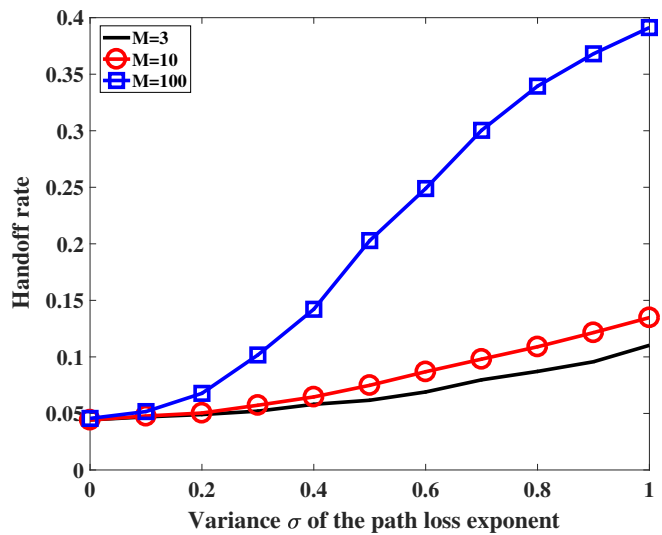}
\par\end{centering}
\caption{\label{fig:The-handoff-rate_sigma}The handoff rate with respect to
the variance $\sigma$ of the path loss exponent, considering different
values of $M$.}
\end{figure}

Fig. \ref{fig:The-handoff-rate_sigma} illustrates the handoff rate
with respect to the variance $\sigma$ of the path loss exponent,
considering different values of $M$. For the analysis of the handoff
rate, the detection interval $t_{d}$ is configured as 1 second, and
the user moving speed $v$ is 5m/s. The handoff rate increases with
increase of the variance $\sigma$ when the value of $M$ is fixed.
The slope of the curve with $M=100$, is much larger than slopes of
the other curves, \textit{i.e.}, the handoff rate with larger $M$
is growing more than the handoff rate with small $M$. The result
shows that the complex anisotropic propagation environment has negative
impact on the handoff performance, and the handoff probability analyzed
based on the isotropic path loss model and the actual handoff probability
have a huge difference in more complex anisotropic propagation environment.

\begin{figure}[tbh]
\subfloat[\label{fig:a}]{\includegraphics[width=8cm]{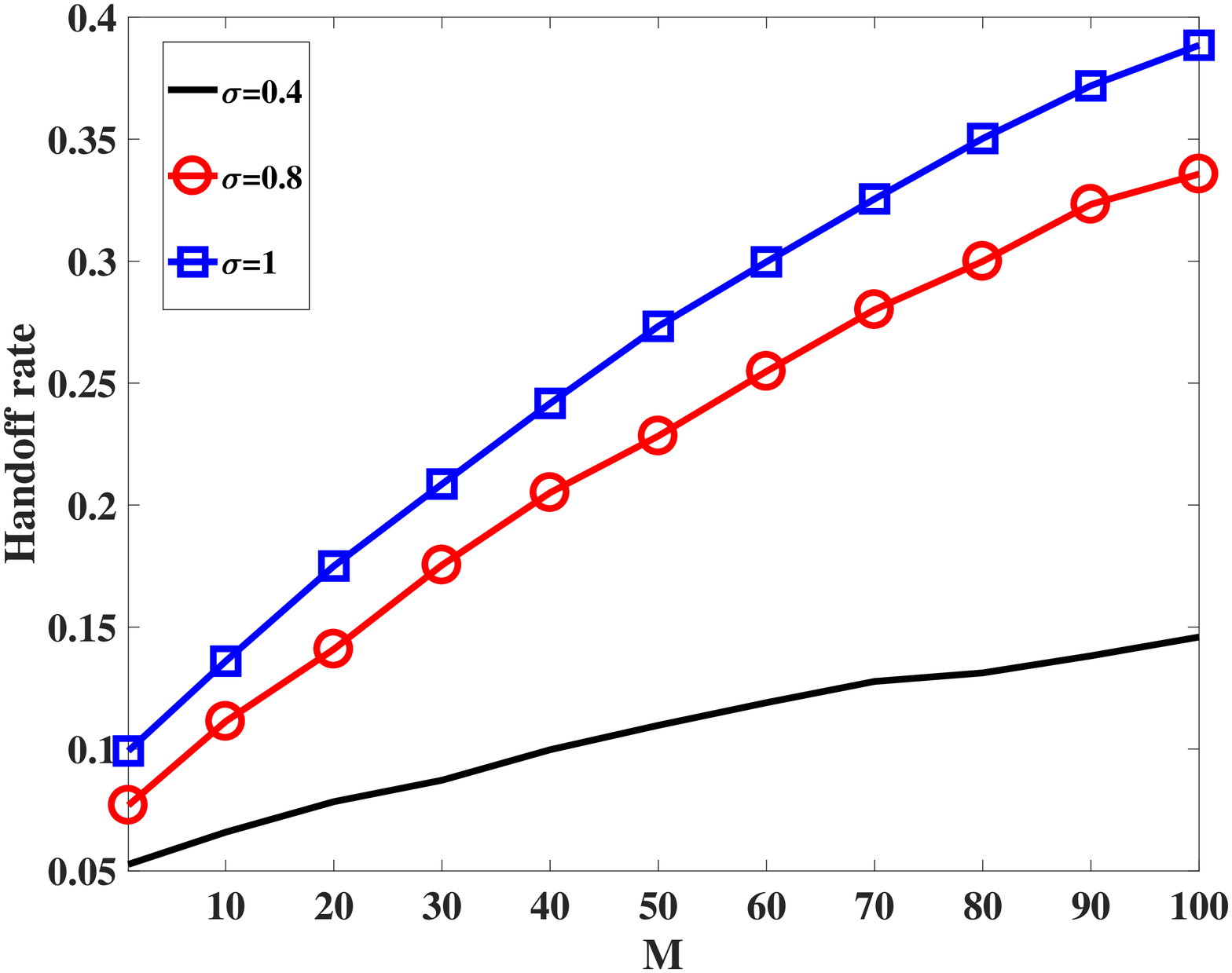}

}\subfloat[\label{fig:b}]{\includegraphics[width=8cm]{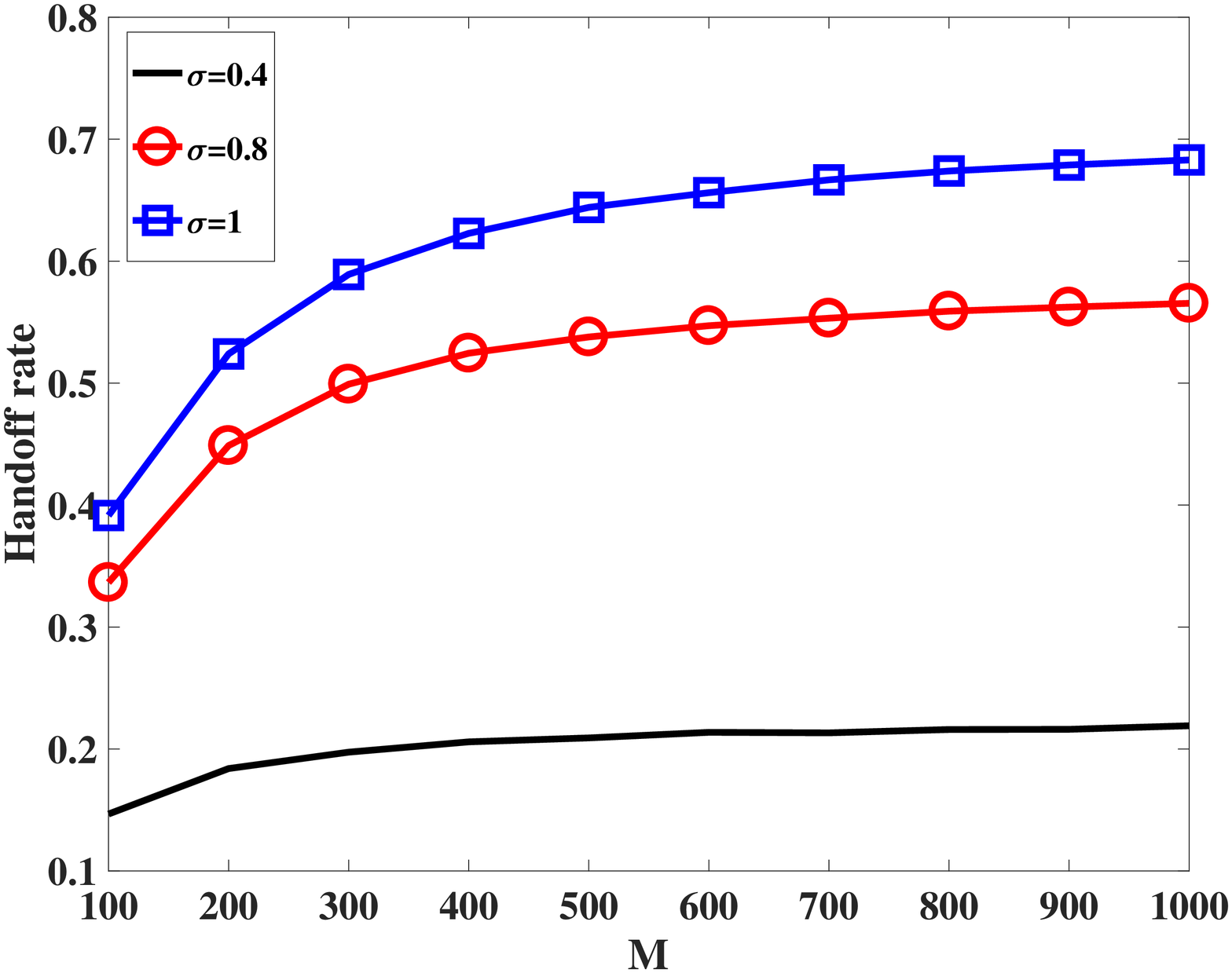}

}

\caption{\label{fig:The-handoff-rate-M}The handoff rate with respect to $M$,
considering different variances $\sigma$ of the path loss exponent.}
\end{figure}

Fig. \ref{fig:The-handoff-rate-M} shows the handoff rate with respect
to $M$, considering different variances $\sigma$ of the path loss
exponent. The handoff rate increases with the increase of $M$, the
number of sections of each SBS. In Fig. \ref{fig:a}, the value of
$M$ is in the range of $\left[10,100\right]$, and the handoff rate
increases rapidly with the increase of $M$. In Fig. \ref{fig:b},
the value of M is in the range of $\left[100,1000\right]$, the increase
of the handoff rate becomes slowdown when $M$ is larger than 400.
The impact of $M$ on the handoff rate becomes stable when $M$ is
larger than 400 which advises that the coverage region of an SBS can
be equally partitioned into 400 sections in practice.

In Fig. \ref{fig:b} the handoff rate with the multi-directional path
loss model has a upper limit which increases with the increase of
the variance. It is because the probability that the path loss exponent
is unchanged during the movement from $l_{0}$ to $l_{1}$ is zero
due to $\frac{2\pi}{M}$ approaching zero when $M$ approaches infinity.
In this case, the path loss exponent of the link at each detection
time is an independent random variable. The limit of the handoff rate
of 5G fractal small cell networks is influenced by the distribution
of the path loss exponent, not influenced by the value of $M$. And
the increase percentage of the handoff rate with the large variance
over that with low variance is more than 100\% when the large variance
is twice as much as the low variance.

\section{Conclusions}

The small cell network is a promising technology for 5G mobile communication
systems to increase the network capacity. Considering the complex
environment in urban scenarios, the anisotropic path loss effect cannot
be ignored any more. A multi-directional path loss model was proposed
to analyze the impact of the anisotropic path loss exponent on performance
in 5G fractal small cell networks. The new coverage probability, association
probability and handoff probability were derived based on the proposed
multi-directional path loss model. Numerical results indicate that
the association probability with short link distance, \textit{e.g.},
50m, decreases obviously with the increase of the effect of anisotropic
path loss in 5G fractal small cell networks. Moreover, it was observed
that the handoff probability with the multi-directional path loss
model is larger than the handoff probability with the isotropic path
loss model. When the coverage area of each SBS is equally partitioned
into 400 sections, the handoff performance can be stably analyzed
with the multi-directional path loss model. When the multi-directional
path loss model with specific parameters for a special scenario is
obtained, the association and handoff probability can be calculated
directly based on the statistic characteristic of the measured parameters,
which can provide straightforward guidance for the deployment of wireless
communication networks.

\appendices{}

\section*{Appendix A: Proof of Lemma 1}

\textit{Proof}: Let $\mathrm{SINR}\left(x_{i}\right)$ be denoted
as $b_{i}$, and $a_{i}=P_{T}G_{i}h_{i}r_{i}^{-\alpha_{i}}$, we have
\begin{eqnarray}
 &  & b_{i}=\frac{\nicefrac{M_{t}M_{r}a_{i}}{G_{i}}}{\underset{j\neq i}{\sum}a_{j}+\sigma_{n}^{2}}=\frac{\nicefrac{M_{t}M_{r}a_{i}}{G_{i}}}{\underset{\Phi}{\sum}a_{j}+\sigma_{n}^{2}-a_{i}}\nonumber \\
 &  & \Longrightarrow\frac{a_{i}}{\underset{\Phi}{\sum}a_{j}+\sigma_{n}^{2}}=\frac{1}{1+\nicefrac{M_{t}M_{r}}{G_{i}b_{i}}}\nonumber \\
 &  & \Longrightarrow\underset{\Phi}{\sum}\frac{1}{1+\nicefrac{M_{t}M_{r}}{G_{i}b_{i}}}+\frac{\sigma_{n}^{2}}{\underset{\Phi}{\sum}a_{j}+\sigma_{n}^{2}}=1.\label{eq:sumeq1}
\end{eqnarray}
Then following the proof of Lemma 1 in \cite{key-31}, (\ref{eq:sumeq1})
could be satisfied if only one of the $b_{i}\,\left(i\in\mathbb{R}^{+}\right)$
is greater than $\frac{M_{t}M_{r}}{m_{t}m_{r}}$, \textit{i.e.}, $\frac{M_{t}M_{r}}{G_{i}b_{i}}<1$,
and $\frac{1}{1+\nicefrac{M_{t}M_{r}}{G_{i}b_{i}}}>\frac{1}{2}$,
because the value of $G_{i}$ is always greater or equal to $m_{t}m_{r}$.
Assumed that two $b_{i}$ are greater than $\frac{M_{t}M_{r}}{m_{t}m_{r}}$,
denoting as $b_{1}$, $b_{2}$. Therefore, $\frac{1}{1+\nicefrac{M_{t}M_{r}}{G_{1}b_{1}}}>\frac{1}{2}$,
$\frac{1}{1+\nicefrac{M_{t}M_{r}}{G_{2}b_{2}}}>\frac{1}{2}$, and
$\frac{1}{1+\nicefrac{M_{t}M_{r}}{G_{1}b_{1}}}+\frac{1}{1+\nicefrac{M_{t}M_{r}}{G_{2}b_{2}}}>1$
which is in contradiction with (\ref{eq:sumeq1}). This contradictory
situation occurs when more $b_{i}$ are greater than $\frac{M_{t}M_{r}}{m_{t}m_{r}}$.
In this case, at most one $b_{i}$ can be greater than $\frac{M_{t}M_{r}}{m_{t}m_{r}}$,
and Lemma 1 is completely proved.

\section*{Appendix B: Proof of Theorem 2}

\textit{Proof}: The coverage probability in a 5G fractal small cell
network with the maximum SINR association can be expressed as
\begin{align}
P_{C}\left(\tau\right) & =\textrm{Pr}\left\{ \underset{x_{i}\in\Phi}{\max}\,\mathrm{SINR}\left(x_{i}\right)>\tau\right\} \nonumber \\
 & =\textrm{Pr}\left\{ \underset{x_{i}\in\Phi}{\bigcup}\frac{P_{T}M_{t}M_{r}h_{i}r_{i}^{-\alpha_{i}}}{I\left(x_{i}\right)+\sigma_{n}^{2}}>\tau\right\} .
\end{align}
Furthermore, according to Lemma 1, when the SINR threshold $\tau>\frac{M_{t}M_{r}}{m_{t}m_{r}}$,
the coverage probability is derived as \textit{
\begin{eqnarray}
 &  & P_{C}\left(\tau\right)=\mathbb{E}\left[\mathbf{1}\left\{ \underset{x_{i}\in\Phi}{\bigcup}\mathrm{SINR}\left(x_{i}\right)>\tau\right\} \right]=\mathbb{E}\left[\underset{x_{i}\in\Phi}{\sum}\mathbf{1}\left\{ \mathrm{SINR}\left(x_{i}\right)>\tau\right\} \right]\nonumber \\
 &  & \overset{\left(a\right)}{=}\lambda\int_{\mathbb{R}^{2}}\Pr\left\{ \frac{P_{T}M_{t}M_{r}h_{i}r_{i}^{-\alpha_{i}}}{I\left(x_{i}\right)+\sigma_{n}^{2}}>\tau\right\} dx_{i}=\lambda\int_{\mathbb{R}^{2}}\Pr\left\{ h_{i}>\frac{\tau r_{i}^{\alpha_{i}}\left(I\left(x_{i}\right)+\sigma_{n}^{2}\right)}{P_{T}M_{t}M_{r}}\right\} dx_{i}\nonumber \\
 &  & \overset{\left(b\right)}{=}\lambda\int_{\mathbb{R}^{2}}\mathbb{E}_{\alpha_{i}}\left[\mathcal{L}_{I}\left(\frac{\tau r_{i}^{\alpha_{i}}}{P_{T}M_{t}M_{r}}\right)e^{-\frac{\tau r_{i}^{\alpha_{i}}\sigma_{n}^{2}}{P_{T}M_{t}M_{r}}}\right]dx_{i},
\end{eqnarray}
}where $\mathbb{E}_{X}\left[\cdot\right]$ denotes the statistical
average with respect to the random variable $X$, $\mathbf{1}\left\{ \cdot\right\} $
denotes the indicator function, step (a) follows from Campbell Mecke
Theorem \cite{key-32}, and step (b) follows from the assumption that
the power gain of the channel fading $h_{i}$ is the exponential distribution
with mean 1 and the definition of Laplace transform. $\mathcal{L}_{I}\left(\cdot\right)$
is the Laplace functional of the aggregated interference power $I\left(x_{i}\right)$.

Considering the aggregated interference power expressed as $I\left(x_{i}\right)=\underset{x_{j}\in\Phi,j\neq i}{\sum}P_{T}G_{j}h_{j}r_{j}^{-\alpha_{j}}$,
the Laplace function of $I\left(x_{i}\right)$ is further derived
as \textit{
\begin{eqnarray}
 &  & \mathcal{L}_{I}\left(s\right)=\mathbb{E}\left[e^{-s\underset{x_{j}\in\Phi,j\neq i}{\sum}P_{T}G_{j}h_{j}r_{j}^{-\alpha_{j}}}\right]=\mathbb{E}_{\Phi}\left[\underset{x_{j}\in\Phi,j\neq i}{\prod}\mathbb{E}_{h_{j},\alpha_{j},G_{j}}\left[e^{-sP_{T}G_{j}h_{j}r_{j}^{-\alpha_{j}}}\right]\right]\nonumber \\
 &  & \overset{\left(c\right)}{=}\exp\left(-\lambda\int_{\mathbb{R}^{2}}\left(1-\mathbb{E}_{h,\alpha,G}\left[e^{-sP_{T}Gh\left|x\right|^{-\alpha}}\right]\right)dx\right)\nonumber \\
 &  & \overset{\left(d\right)}{=}\exp\left(-\frac{\lambda}{2\sqrt{3}\sigma}\int_{\mathbb{R}^{2}}\int_{\alpha}\underset{l=1}{\overset{4}{\sum}}p_{l}\left(1-\frac{1}{1+sP_{T}g_{l}\left|x\right|^{-\alpha}}\right)d\alpha dx\right),
\end{eqnarray}
}where step (c) follows from the probability generating functional
(PGFL) of the PPP \cite{key-32}, which states that $\mathbb{E}_{\Phi}\left[\underset{x\in\Phi}{\prod}f\left(x\right)\right]=\exp\left(-\lambda\int_{\mathbb{R}^{2}}\left(1-f\left(x\right)\right)dx\right)$
for $f\left(x\right)<1$, step (d) is the result of statistical averages
with respect to the antenna gain $G$, the power gain $h$ and the
path loss exponent $\alpha$ with the assumption on the uniform distribution
of the path loss exponent. Let $t=\frac{\left|x\right|^{\alpha}}{sP_{T}g_{l}}$,
we have
\begin{align}
\mathcal{L}_{I}\left(s\right) & =\exp\left(-\frac{\lambda\pi}{\sqrt{3}\sigma}\underset{l=1}{\overset{4}{\sum}}p_{l}\int_{\alpha}\frac{\left(sP_{T}g_{l}\right)^{\frac{2}{\alpha}}}{\alpha}\int_{0}^{\infty}\frac{t^{\frac{2}{\alpha}-1}}{1+t}dtd\alpha\right)\nonumber \\
 & =\exp\left(-\frac{\lambda\pi^{2}}{\sqrt{3}\sigma}\underset{l=1}{\overset{4}{\sum}}p_{l}\int_{\mu-\sqrt{3}\sigma}^{\mu+\sqrt{3}\sigma}\frac{\left(sP_{T}g_{l}\right)^{\frac{2}{\alpha}}}{\alpha}\csc\left(\frac{2\pi}{\alpha}\right)d\alpha\right).\label{eq:interferenceLaplace}
\end{align}
 Let $s=\frac{\tau r_{i}^{\alpha_{i}}}{P_{T}M_{t}M_{r}}$ in (\ref{eq:interferenceLaplace}),
the coverage probability is given as (\ref{eq:covProg-1}), which
completes the proof.

\section*{Appendix C: Proof of Lemma 3}

\textit{Proof}: Let $P$ denote the event that the received signal
power $P_{k}$ is the maximum, and let $S$ denote the event that
$\mathrm{SINR}_{k}$ is the maximum. According to the derivation in
(\ref{eq:sumeq1}) in Appendix A, let $a_{i}=P_{i}$ (the positive
values, not in dB) and $b_{i}=\mathrm{SINR}\left(x_{i}\right)$, we
have
\begin{equation}
\frac{a_{k}}{\underset{\Phi}{\sum}a_{j}+\sigma_{n}^{2}}=\frac{1}{1+\nicefrac{M_{t}M_{r}}{G_{k}b_{k}}}.\label{eq:21}
\end{equation}
The event $P$ is that $a_{k}$ is the maximum, equivalent to that
$G_{k}b_{k}$ is the maximum based on (\ref{eq:21}), which is expressed
as
\begin{align}
G_{k}b_{k} & >G_{i}b_{i}\Rightarrow\frac{b_{k}}{b_{i}}>\frac{G_{i}}{G_{k}}\:x_{i}\in\Phi,\,i\neq k.
\end{align}
The event $S$ is that $b_{k}$ is the maximum, equivalent to that
$\frac{b_{k}}{b_{i}}\geq1$ for $x_{i}\in\Phi,i\neq k$.

When the event $P$ is true, the probability that the event $S$ is
true is expressed as
\begin{equation}
\Pr\left\{ S|P\right\} =\Pr\left\{ b_{k}=\underset{x_{i}\in\Phi}{\max}\,b_{i}|a_{k}=\underset{x_{i}\in\Phi}{\max}\,a_{i}\right\} .
\end{equation}
When $\frac{G_{i}}{G_{k}}\geq1$, $\frac{b_{k}}{b_{i}}>\frac{G_{i}}{G_{k}}\geq1$
denoting that the event $S$ is true. When $\frac{G_{i}}{G_{k}}<1$,
$b_{k}\geq b_{i}$ and $b_{k}<b_{i}$ could be true possibly. Moreover,
let the values of the antenna gain be arranged as the descending order,
$g_{1}\geq g_{2}\geq g_{3}\geq g_{4}$, the conditional probability
that the event $S$ is true is further given as
\begin{align}
\Pr\left\{ S|P\right\}  & \geq\Pr\left\{ \frac{G_{i}}{G_{k}}\geq1,\forall x_{i}\in\Phi,i\neq k\right\} \nonumber \\
 & =\Pr\left\{ G_{k}=g_{1},G_{i}=g_{1}\left(i\neq k\right)\right\} +\Pr\left\{ G_{k}=g_{2},G_{i}\in\left\{ g_{1},g_{2}\right\} \left(i\neq k\right)\right\} \nonumber \\
 & +\Pr\left\{ G_{k}=g_{3},G_{i}\in\left\{ g_{1},g_{2},g_{3}\right\} \left(i\neq k\right)\right\} +\Pr\left\{ G_{k}=g_{4}\right\} \nonumber \\
 & =p_{4}+p_{3}\underset{i\neq k}{\prod}\left(p_{1}+p_{2}+p_{3}\right)+p_{2}\underset{i\neq k}{\prod}\left(p_{1}+p_{2}\right)+p_{1}\underset{i\neq k}{\prod}p_{1}\nonumber \\
 & \overset{\left(a\right)}{=}p_{4}=\left(1-\frac{\phi_{t}}{360}\right)\left(1-\frac{\phi_{r}}{360}\right),
\end{align}
where step (a) is obtained because the three continuous multiplications
are equal to zero when the number of SBSs is infinite.

\section*{Appendix D: Proof of Theorem 4}

\textit{Proof}: The association probability with respect to the distance
$r$ between $u_{0}$ and the desired SBS $\textrm{SBS}^{r}$ is expressed
as
\begin{align}
 & P_{A}\left(r\right)=\Pr\left\{ \mathrm{SINR}\left(r\right)=\underset{x_{i}\in\Phi}{\max}\,\mathrm{SINR}\left(x_{i}\right)\right\} \nonumber \\
 & \approx\Pr\left\{ G_{r}h_{r}r^{-\alpha_{r}}=\underset{x_{i}\in\Phi}{\max}\,G_{i}h_{i}r_{i}^{-\alpha_{i}}\right\} \nonumber \\
 & =\mathbb{E}\left[\underset{x_{i}\in\Phi}{\prod}\mathbb{E}_{G_{i},\alpha_{i}}\left[\Pr\left\{ G_{r}h_{r}r^{-\alpha_{r}}\geq G_{i}h_{i}r_{i}^{-\alpha_{i}}\right\} \right]\right]\nonumber \\
 & =\mathbb{E}\left[\exp\left(-\lambda\int_{\mathbb{R}^{2}}1-\mathbb{E}_{G_{i},\alpha_{i}}\left[\Pr\left\{ h_{i}\leq\frac{G_{r}h_{r}r^{-\alpha_{r}}x^{\alpha_{i}}}{G_{i}}\right\} \right]dx\right)\right]\nonumber \\
 & =\mathbb{E}\left[\exp\left(-\lambda\int_{\mathbb{R}^{2}}\mathbb{E}_{G_{i},\alpha_{i}}\left[\exp\left(-\frac{G_{r}h_{r}r^{-\alpha_{r}}x^{\alpha_{i}}}{G_{i}}\right)\right]dx\right)\right],
\end{align}
where $G_{r}$, $h_{r}$, and $\alpha_{r}$ denote the antenna gain,
power gain from Rayleigh fading, and path loss exponent of the link
between $u_{0}$ and the desired SBS $\textrm{SBS}^{r}$, respectively.
Using the PDFs of the power gain from Rayleigh fading, path loss exponent,
and the probability distribution of antenna gain, the association
probability is further expressed as (\ref{eq:assPro}), which completes
the proof.

\section*{Appendix E: Proof of Theorem 5}

\textit{Proof}: The probability $\Pr\left\{ \mathrm{SINR}^{0}\left(\tilde{r}\right)<\tau_{h}-\tau_{\textrm{hys}}|\alpha_{r}\right\} $
is derived as
\begin{align}
 & \Pr\left\{ \mathrm{SINR}^{0}\left(\tilde{r}\right)<\tau_{h}-\tau_{\textrm{hys}}|\alpha_{r}\right\} \nonumber \\
 & =\mathbb{E}\left[\!\Pr\left\{ \!\tilde{h}_{r}\!<\!\left(\frac{\tau_{h}-\tau_{\textrm{hys}}}{P_{T}M_{t}M_{r}}\right)\!\left(I\left(r\!+\!vt\right)\!+\!\sigma_{n}^{2}\right)\left(r\!+\!vt\right)^{\alpha_{r}}\right\} \right]\nonumber \\
 & \overset{\left(a\right)}{=}1\!-\!\mathbb{E\!}\left[\!\exp\!\left(\!-\!\left(\frac{\tau_{h}-\tau_{\textrm{hys}}}{P_{T}M_{t}M_{r}}\right)\!\left(I\left(r\!+\!vt\right)\!+\!\sigma_{n}^{2}\right)\left(r\!+\!vt\right)^{\alpha_{r}}\right)\right]\nonumber \\
 & =1-\exp\left(-\left(\frac{\tau_{h}-\tau_{\textrm{hys}}}{P_{T}M_{t}M_{r}}\right)\sigma_{n}^{2}\left(r+vt\right)^{\alpha_{r}}\right)\nonumber \\
 & \times\mathcal{L}_{I}\left(\left(\frac{\tau_{h}-\tau_{\textrm{hys}}}{P_{T}M_{t}M_{r}}\right)\left(r+vt\right)^{\alpha_{r}}\right),
\end{align}
where (a) is obtained by the PDF of the power gain $\tilde{h}_{r}$
at the location $l_{1}$ which is \textit{i.i.d.} from the power gain
$h_{r}$ at the location $l_{0}$. Let $s=\left(\frac{\tau_{h}-\tau_{\textrm{hys}}}{P_{T}M_{t}M_{r}}\right)\left(r+vt\right)^{\alpha_{r}}$
in (\ref{eq:interferenceLaplace}),
\begin{align}
 & \Pr\left\{ \mathrm{SINR}^{0}\left(\tilde{r}\right)<\tau_{h}-\tau_{\textrm{hys}}|\alpha_{r}\right\} \nonumber \\
 & =1-\exp\left(-\left(\frac{\tau_{h}-\tau_{\textrm{hys}}}{P_{T}M_{t}M_{r}}\right)\sigma_{n}^{2}\left(r+vt\right)^{\alpha_{r}}\right)\nonumber \\
 & \times\!\exp\left(\!-\frac{\lambda\pi^{2}}{\sqrt{3}\sigma}\!\underset{l=1}{\overset{4}{\sum}}p_{l}\!\int_{\alpha}\!\frac{\left(\frac{\left(\tau_{h}-\tau_{\textrm{hys}}\right)\left(r+vt\right)^{\alpha_{r}}g_{l}}{M_{t}M_{r}}\right)^{\frac{2}{\alpha}}\!\csc\left(\frac{2\pi}{\alpha}\right)}{\alpha}\!d\alpha\right).\label{eq:cond1}
\end{align}
Substitute (\ref{eq:pa_r}) and (\ref{eq:cond1}) into (\ref{eq:hpro1}),
the handoff probability with $\varphi=\Theta^{r}$ is given as (\ref{eq:handoff1}),
which completes the proof.

\end{document}